\documentclass[aps,prl,reprint,amssymb,showpacs,superscriptaddress,notitlepage,onecolumn]{revtex4-2}
\usepackage{bm}
\usepackage{graphicx}
\usepackage{hyperref}
\usepackage{dcolumn}
\usepackage{braket}
\usepackage{natbib}
\usepackage{amsmath}
\usepackage{color}
\usepackage{mathtools,amssymb}
\usepackage{soul}
\usepackage{float}
\usepackage{amssymb}
\usepackage{esint}
\usepackage{mathtools}
\usepackage{physics}
\usepackage{makecell}
\usepackage{multirow}
\usepackage{array}
\usepackage{geometry}
\usepackage{url}
\usepackage{adjustbox}
\usepackage{minted}
\usepackage{rotating}
\usepackage{listings}
\usepackage{threeparttable,booktabs}
\usepackage{fontawesome}
\usepackage{comment}
\usepackage{cleveref}
\usepackage{cuted}
\usepackage{tikz}
\usepackage{titletoc} 

\definecolor{beaver}{rgb}{0.62, 0.51, 0.44}
\definecolor{bole}{rgb}{0.47, 0.27, 0.23}
\definecolor{brown(traditional)}{rgb}{0.59, 0.29, 0.0}
\definecolor{burntorange}{rgb}{0.8, 0.33, 0.0}
\usepackage{svg}

\newcommand{\tx}{{\times}}
\makeatletter
\newenvironment{supplement}{%
  \setcounter{secnumdepth}{3}%
  \setcounter{section}{0}%
  \def\@seccntformat##1{\csname the##1\endcsname:\ }%
}{%
  \def\@seccntformat##1{\csname the##1\endcsname.\ }%
}
\makeatother

\newcommand{\SMsec}[1]{Sec.~\ref{#1}}

\begin{document}

\title{Quantum Metamorphosis: Programmable Emergence and the Breakdown of Bulk-Edge Dichotomy in Multiscale Systems}

\author{Mahmoud Jalali Mehrabad$^{\ddagger }$}\email{mjalalim@mit.edu}
\affiliation{Research Laboratory of Electronics, Massachusetts Institute of Technology, Cambridge, Massachusetts 02139, USA}

\author{Alireza Parhizkar$^{\ddagger }$}\email{alpa@umd.edu}
\affiliation{Joint Quantum Institute, University of Maryland and National Institute of Standards and Technology,
College Park, MD 20742, USA}

\author{Lida Xu}
\altaffiliation{Equally contributing authors}
\affiliation{Joint Quantum Institute, University of Maryland and National Institute of Standards and Technology,
College Park, MD 20742, USA}

\author{Gregory Moille}
\affiliation{Joint Quantum Institute, University of Maryland and National Institute of Standards and Technology,
College Park, MD 20742, USA}

\author{Avik Dutt}
\affiliation{Department of Mechanical Engineering, and Institute for Physical Science and Technology, University of Maryland, College Park, Maryland 20742, USA}

\author{Dirk Englund}
\affiliation{Research Laboratory of Electronics, Massachusetts Institute of Technology, Cambridge, Massachusetts 02139, USA}

\author{Kartik Srinivasan}
\affiliation{Joint Quantum Institute, University of Maryland and National Institute of Standards and Technology,
College Park, MD 20742, USA}

\author{Daniel Leykam}
\affiliation{Science, Mathematics and Technology Cluster, Singapore University of Technology and Design, Singapore}

\author{Mohammad Hafezi}\email{hafezi@umd.edu}
\affiliation{Joint Quantum Institute, University of Maryland and National Institute of Standards and Technology,
College Park, MD 20742, USA}

\begin{abstract}
Multiscale synergy---the interplay of a system's distinct characteristic length, time, and energy scales---is becoming a unifying thread across many contemporary branches of science. Ranging from moir\'{e} and super-moir\'{e} materials~\cite{craig2024local,adak2024tunable,nuckolls2024microscopic}  and cold atoms~\cite{celi2014synthetic,cooper2019topological} to DNA-templated superlattices~\cite{jing2025dna} and nested photonic networks~\cite{flower2024observation}, multiscale synergy produces behaviors not obtainable at any single scale alone~\cite{anderson_more_1972,chalker1988percolation}. Yet a general framework that \emph{programs} cross-scale interplay to steer spectra, transport, and topology has been missing. Here, we elevate multiscale synergy from a byproduct to a general design principle for emergent phenomena. Specifically, we introduce a scale-programmable framework for hierarchically nested lattices (HNLs) that can host quantum metamorphosis (\emph{\textbf{Q}u\textbf{M}orph)}---a continuous evolution between system-dependent features governed by a dimensionless tunable parameter $\alpha$ (the relative hopping). To exemplify, we show an HNL, in which as \(\alpha\) changes, the spectrum metamorphoses from integer quantum Hall–like to anomalous quantum Hall–like, passing through a cocoon regime with proliferating mini-gaps. This multiscale mixing yields multiple novel phenomena, including hybrid edge–bulk states, scale-dependent topology, topologically embedded flat bands, and isolated edge bands. We propose a feasible photonic implementation using commercially available coupled-resonator arrays, outline spatial–spectral signatures to map QuMorph, and explore applications for multi-timescale nonlinear optics. Our work establishes a scalable and programmable paradigm for engineering multiscale emergent phenomena.

\end{abstract}

\maketitle
\section{Introduction}

Multiscale structures are transitioning from complications to powerful resources across physics and engineering. Moir\'{e} and super-moir\'{e} materials~\cite{cao2018unconventional,cao2018correlated,balents2019general,bistritzer2011moire,parhizkar2022strained,craig2024local,oudich2024engineered,du2023moire,adak2024tunable,nuckolls2024microscopic,he2025strongly} are prime examples in which tiny angular or lattice mismatches inflate real-space periods and shrink Brillouin zones, reshaping dispersion and interaction landscapes. Moreover, in ultracold atoms, internal states can emulate extra scale-independent synthetic dimensions with engineered gauge fields, enabling higher-dimensional band topology~\cite{celi2014synthetic,cooper2019topological}. Another example is bottom-up molecular platforms---e.g., DNA-templated lattices---that assemble programmable nanometer-scale registries into larger superstructures~\cite{jing2025dna}. Photonic resonator arrays are another emergent platform that maps geometric nesting into multiple, widely separated dynamical timescales~\cite{mittal2021topological,
huang2024hyperbolic,hashemi2024floquet,tusnin2023nonlinear,hashemi2025reconfigurable,flower2024observation,xu2025chip,pang2025versatile,mehrabad2025multi,jalali2023topological}. Despite their differences, these systems share a common mechanism: small, controllable offsets at one scale seed qualitatively new behavior at another.

What is still lacking is (i) a general scale-programmable way to \emph{steer} this behavior by design—treating scale separation not as a platform-specific detail but as the organizing variable; (ii) equally missing is a framework that tracks how multiple scales compete, blurring the clean bulk–edge division assumed by single-scale theories: as the cross-scale ratios shift, energy spectra can reorganize, gaps open and close, edge bands appear and disappear, and transport pathways reconfigure. A framework with a minimal number of dimensionless control knobs, free of fragile fine-tuning, would enable purposeful navigation across regimes that are usually studied in isolation.

\begin{figure*}[t]
    \centering
    \includegraphics[width=0.73\textwidth]{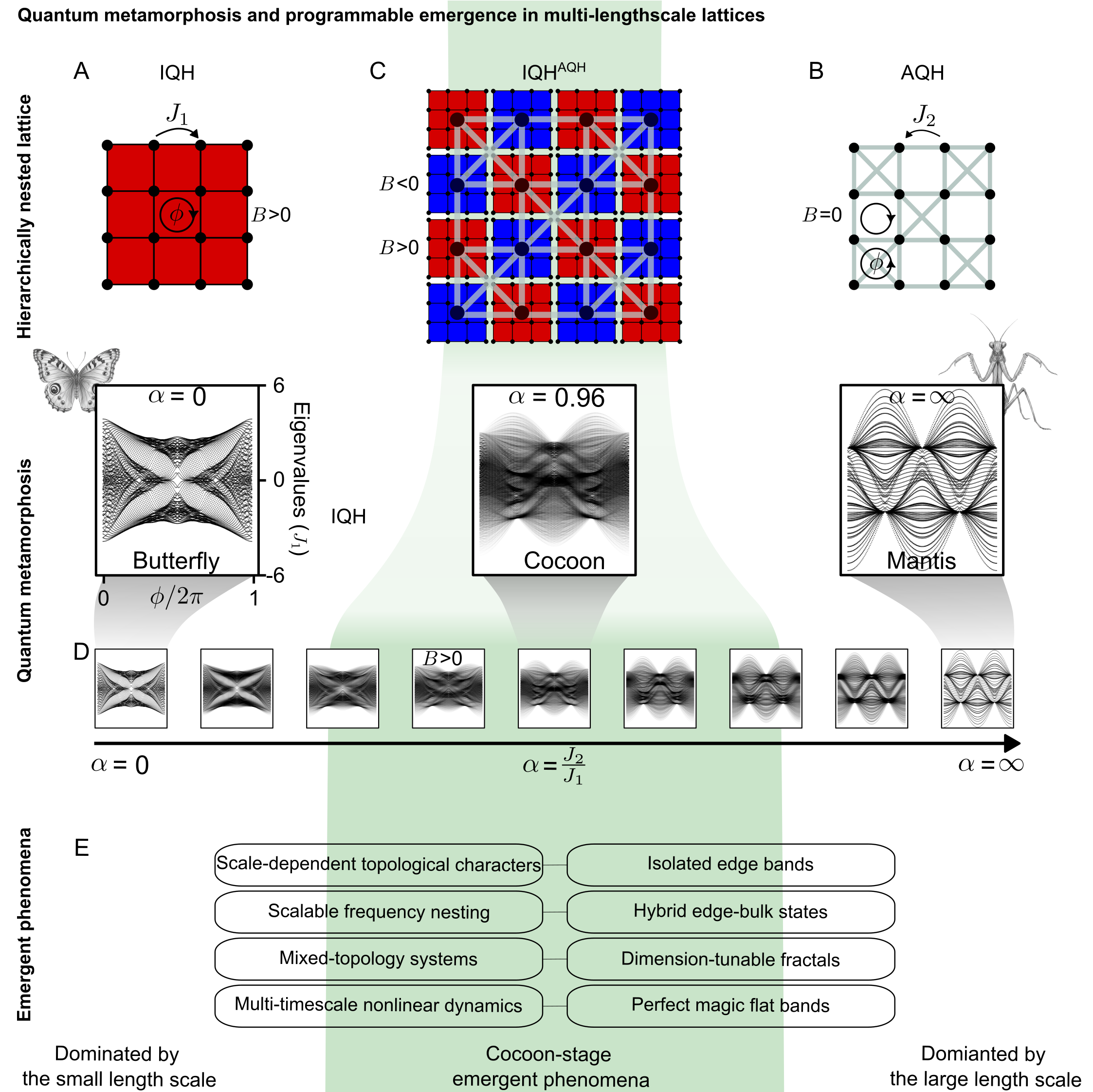}
    \caption{The general notion of quantum metamorphosis (QuMorph) in hierarchically nested lattices (HNLs). (A) A $4 \tx 4$ first-order nested square lattice with flux $\phi$ threading each plaquette, yielding an integer quantum Hall-like (IQH-like) lattice, hosting the renowned Hofstadter's \emph{butterfly} as its energy-flux spectrum. The edge states appear as the sparse levels resembling a butterfly. (B) An anomalous quantum Hall-like (AQH-like) $4\tx 4$ first-order nested square lattice, where the total flux through each unit cell is zero. The corresponding energy spectrum of this system resembles that of a \emph{mantis}, with the edge states similarly appearing as sparse levels. (C) A second-order $4\tx 4^{4\tx 4}$ HNL architecture, in which individual copies of small IQH lattices (with alternating magnetic fields $\pm{B}$, marked by red/blue) serve as the sites of the larger AQH lattice, forming a $\text{IQH}^\text{AQH}$ HNL. Hopping rate between the sites within an IQH lattice (between copies of IQH lattices in the super AQH lattice) is set by $J_1$ ($J_2$). (D) Tuning the relative hopping $\alpha = J_2/J_1$ from $0$ to $\infty$ transforms the effective system from the IQH to the AQH phase, yielding the metamorphosis of flux-energy spectrum from the butterfly into the mantis. The crossover is a \emph{cocoon} stage that lies in the middle. (E) Representative emergent phenomena at the cocoon stage: (left) programmable and (right) designed emergent phenomena.}
    \label{Fig:Morph}
\end{figure*}

We address both gaps by formulating a scale-controllable hierarchically nested lattice (HNL) design framework that exhibits the emergence of \emph{quantum metamorphosis (QuMorph)}---a continuous transformation between scale-dependent phases governed by one dimensionless ratio $\alpha$ that sets the relative weight of the hopping rates denoted by $J_1$ and $J_2$, respectively. As a prominent example, we demonstrate the morphing of the energy-flux spectrum from an integer quantum Hall (IQH-like) to an anomalous quantum Hall (AQH-like) system. By tuning \(\alpha\) in the system, the spectrum undergoes \emph{QuMorph}: the canonical Hofstadter \emph{butterfly}~\cite{hofstadter1976energy} (IQH-like) continuously reshapes as inter-order coupling increases, passing through a \emph{cocoon} regime where competing scales generate a dense spectrum of mini-gaps and hybrid edge–bulk bands. Through controlled gap closings/reopenings, new Chern numbers emerge and reorganize, and magic perfectly-flat bands appear. Beyond the crossover, the spectral pattern settles into an AQH-like \emph{mantis}: lobed bands framed by isolated edge manifolds and a narrowed, nearly dispersion-less central spine. This metamorphosis is driven purely by scale mixing and is programmed solely by \(\alpha\). Our HNL architecture and the programmable emergence of QuMorph are summarized in Fig.~\ref{Fig:Morph}. Our general scheme is platform-agnostic wherever tunneling can be engineered. For example, various tunneling Hamiltonians can be optically engineered for ultracold atoms on optical lattices~\cite{bloch2008many}. More intriguingly, a broad range of artificial gauge fields can be implemented, where the light-assisted tunneling provides the required gauge phases, both in real and synthetic dimensions~\cite{dalibard2011colloquium,galitski2019artificial}. Moreover, recent advances in superconducting circuit-QED platforms have enabled engineering a wide range of lattice geometries and coupling waveguides~\cite{schmidt2013circuit, carusotto2020photonic}, e.g., chiral and synthetic gauge fields~\cite{roushan2017chiral,rosen2024synthetic}, metamaterial structures~\cite{mirhosseini2018superconducting}, and hyperbolic lattices~\cite{kollar2019hyperbolic}. In this work, our exemplary system of choice is arrays of commercially available ring resonators, where signatures of multi-scale physics have been recently demonstrated in the near-infrared and visible bands~\cite{flower2024observation,xu2025chip,mehrabad2025multi}. We prototypically reveal several exotic and unprecedented physical consequences of QuMorph for integrated photonics. Specifically, we propose realizable photonic circuits comprised of coupled ring-resonators and show the programmable emergence of several striking phenomena, as well as novel device functionalities, including:

\begin{enumerate}
    \item[(i)] Hybrid edge-bulk states
    \item[(ii)] Scale-dependent topological characters
    \item[(iii)] Perfect topologically embedded flat bands
    \item[(iv)] Isolated edge bands
    \item[(v)] Dimension-tunable fractals
    \item[(vi)] Mixed-topology states
    \item[(vii)] Scalable frequency nesting
    \item[(viii)] Multi-timescale nonlinear dynamics
\end{enumerate}

Finally, we comprehensively discuss the details for near-term commercially available fabrication, as well as excitation and detection techniques for experimental spatial-spectral signatures of QuMorph (Supplementary Materials (SM)~\SMsec{sm:exp}). 


\section{Concept}
\subsection{Metamorphosis}

We introduce the general theoretical description of QuMorph by presenting the simplest single-particle species Hamiltonian that exhibits metamorphosis, given by
\begin{equation} \label{eq:AllToAll}
    H = \frac{1}{1+\alpha} (I_f\otimes H_i) + \frac{\alpha}{1+\alpha} (H_f \otimes I_i) + \text{non-commuting terms} \, ,
\end{equation}
where $I_{f,i}$ denote identity matrices matching $H_{f,i}$ in dimension and the prefactors are normalized so that the two terms form a convex combination (weights summing to unity) while $\alpha$ equals the ratio of the weights. To streamline the following discussion, we `trivialize' Eq.~\eqref{eq:AllToAll} by disregarding the non-commuting terms momentarily.
However, even in this limit the single-species constraint preserves nontriviality by enforcing connectivity of $i$- and $f$-sectors (SM~\SMsec{sm:stat}).
Tuning $\alpha$ from $0$ to $\infty$ continuously evolves the Hamiltonian $H$ from $I_f\otimes H_i$ to $H_f\otimes I_i$, carrying the spectral distribution of $H$ from that of (duplicates of) $H_i$ to (duplicates of) $H_f$. When $H_i$ and $H_f$ describe lattice systems, $H$ describes a nested (multiscale) lattice: each site of the $H_f$ lattice is replaced by a copy of the $H_i$ lattice (Fig.~\ref{Fig:Morph}A-C). Couplings between the $H_i$ lattices are set by $H_f$. For example, if an $H_i$ lattice sits at the position $\vec{r}_f$ in the $H_f$ lattice and its neighbor at $\vec{r}'_f$, then each site of the $H_i$ lattice at $\vec{r}_f$ is connected to the corresponding site of the $H_i$ lattice at $\vec{r}'_f$, with the hopping amplitude given by $\bra{\vec{r}_f} H_f \ket{\smash{\vec{r}'_f}}$. Such a second-order nested lattice (a lattice of lattices, i.e., Fig.~\ref{Fig:Morph}C) is naturally labeled by pairs of coordinates: $\vec{r}_f$ and $\vec{r}_i$ for positions in $H_f$ and $H_i$ lattices. With this notation, matrix elements take the following form:
\begin{equation}
    \bra{\vec{r}_f,\vec{r}_i} H \ket{\smash{\vec{r}'_f,\vec{r}'_i}} =\delta_{\vec{r}_f,\vec{r'}_f}\frac{1}{1+\alpha} \bra{\vec{r}_i} H_i \ket{\vec{r}'_i} + \delta_{\vec{r}_i,\vec{r'}_i}\ \frac{\alpha}{1+\alpha} \bra{\vec{r
    }_f} H_f \ket{\smash{\vec{r}'_f}} \, ,
\end{equation}
where $\delta_{a,b}$ is the Kronecker delta.

As a representative case, let $H_i$ describe a lattice subject to a perpendicular uniform magnetic field---an IQH-like system---while $H_f$ describes an AQH-like system (see~\SMsec{sm:qumorphall} and Supplementary Movies 1-3 for other configurations). As the flux through each plaquette varies, the energy-flux spectrum of $H_i$ forms a Hofstadter's butterfly~\cite{hofstadter1976energy}. Accordingly, at $\alpha = 0$ the nested system $H$ inherits a highly degenerate Hofstadter's butterfly. As $\alpha$ increases, the spectrum passes through a cocoon stage near $\alpha=1$ and then completes the metamorphosis, reorganizing into a \emph{mantis} pattern. This QuMorph process is shown in Fig.~\ref{Fig:Morph}D, where $\alpha$ is the ratio of the intra-lattice hopping rates (that of $H_f$ over $H_i$). The cocoon stage is robustly defined as the $\alpha = \alpha_c$ at which the energy spectrum of $H$ is distributed most uniformly (e.g., where the sum of squared level spacings, $\sum \Delta E^2$, is minimum), and is loosely used to refer to $\alpha_c$ and its neighborhood.

At $\alpha=0$ and $\alpha=\infty$, the topology of the entire system can be characterized by a single \emph{first} Chern number. By contrast, in the cocoon stage, a richer topological character emerges. At $\alpha = 1$, the four coordinates $\vec{r}_{i,f}$ appear on an equal footing and the system can be viewed as effectively $(4+1)$-dimensional~\footnote{For simplicity, take $H_i$ and $H_f$ both to describe $(2+1)$-dimensional quantum Hall systems here.}. In four dimensions, a \emph{second} Chern number can be assigned to the system as per the modern theory of polarization, given by~\cite{zhang2001four,kraus2013four-dimensional}
\begin{equation} \label{2ndC}
    C^{(2)} \equiv \frac{1}{32\pi^2}\int d^2 k_i d^2k_f \, \epsilon^{\mu\nu\rho\sigma}\mathcal{F}_{\mu\nu}\mathcal{F}_{\rho\sigma} \, ,
\end{equation}
where $\mathcal{F}_{\mu\nu}$ is the Berry curvature, $\epsilon^{\mu\nu\rho\sigma}$ is the totally anti-symmetric tensor, and the integral is over the entire four-dimensional Brillouin zone~\footnote{We note that the Einstein summation rule is assumed throughout the paper}. Thus, tuning $\alpha$ into the cocoon stage leads to the emergence of topological characters that were previously absent in the system; consequently, the topological invariants classifying the system before and after the metamorphosis need not share the same character.

A similar line of reasoning yields a general criterion for the existence of flat bands. If a flat band exists in the system, on the flat band, the time dimension can be neglected (i.e., the system can be treated as non-propagating), leading to a dimensional reduction from $(4+1)$ to $(4+0)$. In four dimensions, however, a chiral anomaly arises for both Abelian and non-Abelian theories. This anomaly obstructs flat band realization except in specific configurations where it cancels. The cancellation condition furnishes a flat band criterion~\cite{parhizkar2024generic,parhizkar2024localizing}. As with higher Chern numbers, our minimal construction thus can manifest higher-dimensional anomalies (\SMsec{sm:flatband}).

The Hamiltonian~\eqref{eq:AllToAll}, does not attribute any notion of hierarchy to $H_{i,f}$ lattices: $H_{i,f}$ are not distinguishable as the ``inner'' or the ``outer'' lattice. One can presume $H_i$ is nested in $H_f$ or vice versa. Since all corresponding sites are identically coupled, $H_i$ and $H_f$ are treated in the same manner. The hierarchy can be introduced by simply breaking the hopping symmetry through substituting $I_i$ with $\tilde{I}_i \neq I_i$,
\begin{equation} \label{eq:SomeToSome}
    H = \frac{1}{1+\alpha} (I_f\otimes H_i) + \frac{\alpha}{1+\alpha} (H_f \otimes \tilde{I}_i) \, .
\end{equation}
By this choice, $H_i$ and $H_f$ are no longer treated in the same manner, and the connection between $H_i$ copies becomes site dependent. For example, setting $\tilde{I}_i = o I_{\text{Bulk}} + O I_{\text{Edge}}$ preferentially strengthens edge–edge links relative to bulk–bulk links when choosing $O > o$.

When $H_{i,f}$ are taken to be finite-size topological lattices with a non-trivial Chern number, bulk–boundary correspondence ensures edge and bulk states in their spectra~\cite{hatsugai1993chern,vaidya2023response}. Consequently, the spectrum of $H$ hosts hybrid edge–bulk states: $\text{bulk}^\text{bulk}$, $\text{bulk}^\text{edge}$, $\text{edge}^\text{bulk}$, and $\text{edge}^\text{edge}$—where the base label denotes the building-block (smallest) lattice in our notation. For instance, $\text{edge}^\text{bulk}$ is edge-like within $H_i$ but bulk-like within $H_f$. We use this notation throughout this work.

The generalization to higher-order nesting takes the form
\begin{equation}
    H = \sum_{j} \alpha_j \left( \bigotimes_{i<j} I_i \right) \otimes H_j \otimes \left( \bigotimes_{k>j} I_k \right) \, ,
\end{equation}
where $H_j$ is the Hamiltonian of the $j$th nested layer with $\alpha_j$ being its energy scale.
Moreover, Hamiltonians of the form
\begin{equation} \label{eq:PHPI}
    H = \frac{1}{1+\alpha} (I_f\otimes H_i) + \frac{\alpha}{1+\alpha} \left(\sum_P P H_f P \otimes \tilde{I}^P_i\right) \, ,
\end{equation}
where $P$ denotes a set of projectors, can be approximated as Eq.~\eqref{eq:SomeToSome} by exploiting system symmetries.
Prominent examples describable by these Hamiltonians include moir\'{e} systems, where supercells are described by $H_i$ and are coupled via $H_f$ with coupling localized to supercell interfaces by an appropriate $\tilde{I}^P_i$. Their low-energy continuum descriptions (e.g., bilayer graphene) can likewise be captured by patched IQH-like physics (cf. Refs.~\cite{parhizkar2024zero,parhizkar2024generic}).

\begin{figure*}[t]
    \centering
    \includegraphics[width=0.7 \textwidth]{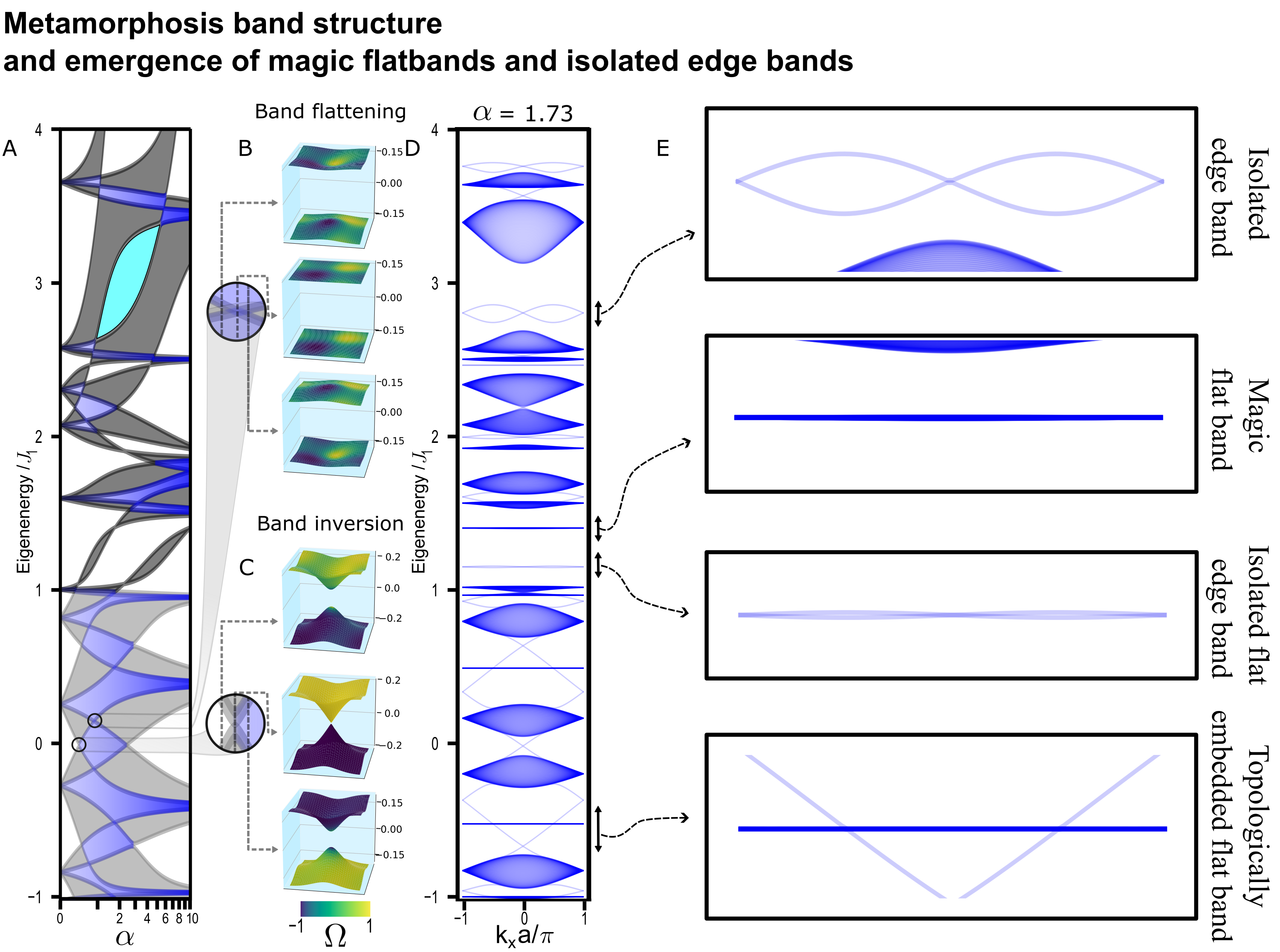}
    \caption{For an $\text{AQH}^\text{AQH}$: (A) Energy bands of a $4\tx 4^{\infty\tx\infty}$ HNL as a function of $\alpha$. The gray shades indicate the bulk modes of the entire HNL. White regions denote trivial gaps; blue regions are topological with $|C|=1$. The light (dark) gray shade represents the $\text{edge}^\text{bulk}$ ($\text{bulk}^\text{bulk}$) states. The cyan shade is an example of a gap with residue topology, within which an isolated (detached) edge band exists. The isolated edge band is created by two band crossings, resulting from higher orbital mixing. After the first crossing, the gap carries $|C|=1$; by the bulk-boundary correspondence, the separated bulk bands must be connected by an edge band. A second crossing at a different symmetry point restores $C=0$, lifting the connectivity requirement and leaving a detached edge band. (B,C) Magnified examples (circled) of multiple instances of band crossing and band flattening observed in (A), plotted along the two-dimensional energy-momentum dispersions. (B) shows the flattening of two adjacent bands; (C) shows a band crossing. Note that the flattenings are perfect, e.g., at $\alpha \approx 0.87$ a perfect flat band sits at $E \approx \pm 0.15 J_1$. The color encodes the point-wise Berry curvature $\Omega(\vec k)$, normalized to $[-1,1]$ within each band. (D) A cross-section of (A) at $\alpha=1.73$, within the cocoon regime where orbital mixing yields emergent phenomena. (E) From top: isolated edge bands; magic perfect flat bands; isolated flat edge bands; and topologically embedded perfect flat bands. Note that in the latter, the flat band is surrounded by topological gaps and thus threaded by edge bands. Apart from these novel behaviors, the spectrum in (D) includes $\text{bulk/edge}^\text{bulk/edge}$ bands; Dirac cones; and connected (regular) edge bands, all in one instance of $\alpha$.
    }
    \label{Fig:Band}
\end{figure*}

A broad family of choices for $\tilde{I}^P_i$ can yield metamorphoses and mixed topologies. Below we consider the simplest case of $\tilde{I}^P_i$, by which the $H_i$ AQH-like lattices are connected at their corners via an AQH-like $H_f$ (see~\SMsec{sm:latham} for the explicit lattice structure). The band structure of such a configuration is shown in Fig.~\ref{Fig:Band}, demonstrating multiple emergent phenomena, analyzed below (see~\SMsec{sm:bandev} for complete band evolution during QuMorph). 

\begin{figure*}[t]
    \centering
    \includegraphics[width=0.80
    \textwidth]{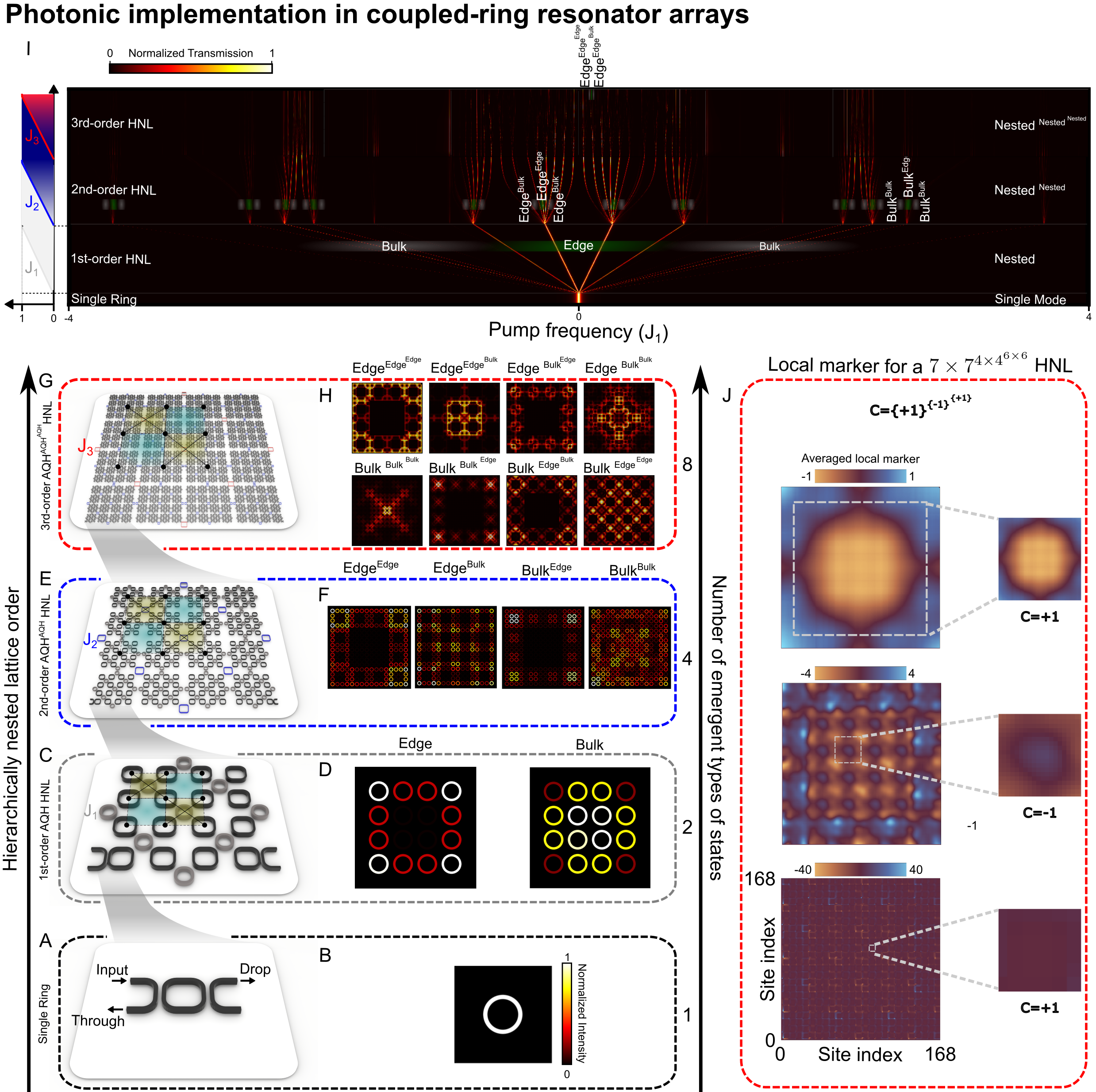}
    \caption{(A-I) Bottom-up scalable hierarchical construction of HNLs in coupled photonic ring-resonator architectures, with spatial-spectral (field intensity profiles and drop-port transmission spectra) spectroscopic signatures for realizing QuMorph in HNL devices in an add-drop configuration (here only for the AQH-type HNLs). Different types of hybrid modes hosted in each HNL are labeled. In (I), resonant-frequency nesting scales clearly with increasing nesting order (only a single longitudinal mode of the single-ring resonator is considered). Frequency bands for each mode type are highlighted and labeled with gray and green, distinguishing the bulk and edge-type bands, respectively. (J) The coarse-grained local Chern marker at three different scales depicted for an $\text{AQH}^{\text{AQH}^\text{AQH}}$ HNL of size $7\tx 7^{4\tx 4^{6\tx 6}}$. Color encodes the coarse-grained Chern marker averaged around each point. The bulk Chern value of the HNL flips as the coarse-graining scale increases from bottom to top.
    }
    \label{Fig:implement}
\end{figure*}
%


\section{Emergence of hybrid edge-bulk wave-functions}
\label{sec:BulkEdge}

QuMorph in the HNLs leads to the emergence of a hybrid combination of edge-bulk states beyond the conventional bulk-edge correspondence. As shown in Fig.~\ref{Fig:implement}, these systems introduce a lattice-order degree of freedom where the edge or bulk character of an eigenstate is set independently at each order (\SMsec{sm:eigen} and~\SMsec{sm:large}). We illustrate this focusing on the simplest case: a second-order HNL in which the inner (\(H_i\)) and outer (\(H_f\)) lattices are finite AQH systems. In our notation, this is an \(\text{AQH}^{\text{AQH}}\) lattice. In this example, the hoppings are set by two parameters: \(J_1\) within \(H_i\), and \(J_2\) between the corners of neighboring \(H_i\) copies that form the \(H_f\) lattice (cf. Fig.~\ref{Fig:implement}G and Fig.~\ref{Fig:Morph}).
For $J_2=0$, the system reduces to isolated copies of \(H_i\). For an infinite \(H_i\) lattice with a gapped bulk, the first positive-energy band onsets near \(E \approx J_1 \). Thus, for a finite \(H_i\), states with \(-J_1 \lesssim E \lesssim J_1\) are edge states of the inner lattice. Outside this energy range, bulk states appear. In Fig.~\ref{Fig:Band}, this window is shaded light gray to indicate edge states of the inner lattice. Dark gray, on the other hand, denotes the bulk states of the inner lattice. As \(J_2\) increases from zero, the corner-to-corner coupling favorably links edge modes, so edge states hybridize more strongly than bulk states. Consequently, edge bands split more rapidly (with $J_2$ increasing) than bulk bands (Fig.~\ref{Fig:Band}). The outer lattice is also AQH, with edge states appearing within a window of \(\pm J_2\). Thus, as the \(H_i\) orbitals bifurcate, all the newly opened gaps are topological with unit Chern number. Accordingly, these gaps host the edge states of the outer lattice: \(\text{edge}^{\text{edge}}\) for \(-J_1<E<J_1\) and \(\text{edge}^{\text{bulk}}\) for \(|E|>J_1\). Outside the gaps, states are \(\text{bulk}^{\text{edge}}\) for \(-J_1<E<J_1\) and \(\text{bulk}^{\text{bulk}}\) otherwise. As \(J_2\) grows, the bifurcations widen and, near \(\alpha=J_2/J_1\approx 1\), intersect at the \(\Gamma\) point (a level crossing). Moreover, at certain values of \(\alpha\), the bands flatten at all momenta. We analyze these flat bands in~\SMsec{sm:flatband} and Supplementary Movie 4.
Taking \(H_f\) to be infinitely large, we impose Bloch periodicity with a unit cell consisting of two \(H_i\) copies. The resulting band structure is shown in Fig.~\ref{Fig:Band}. For example, if \(H_i\) is a \(4\tx 4\) AQH lattice, the composite unit cell hosts 32 orbitals, 16 below and 16 above zero energy.
We calculate the Chern number via the conventional method of integrating the Berry curvature over isolated bands~\cite{thouless1982quantized,kohmoto1985topological,niu1985quantized,berry1984quantal,xiao2010berry}. Topological gaps (with non-trivial Chern number) are shaded blue. The edge states of the finite $H_f$ lattice lie within these gaps. However, this band-based calculation is blind to the topology of the inner lattice \(H_i\): the Chern number is defined per Brillouin zone, which is too small to resolve \(H_i\)’s topological features---a scale-mismatch problem. Nevertheless, we can bypass the problem and assign a topological character to all scales using a coarse-grained local Chern marker, the subject of the next section.

As shown in the field profiles in Fig.~\ref{Fig:implement}A-H, another intriguing phenomenon in our HNLs is the emergence of highly tunable fractal patterns. A detailed study of such fractals is presented in~\SMsec{sm:fract} and Supplementary Movie 5.
\section{{scale-dependent topological characters}}
\label{sec:LocalChern}

The Chern number is the Brillouin-zone integral of the Berry curvature $\Omega(k)$, which can be defined as the trace of a curvature operator $\hat{\Omega}$,
\begin{equation}
    C^{(1)}= \frac{1}{2\pi}\int_{\text{BZ}} \Omega(k) d^2k = \frac{1}{N}\Tr_r \hat{\Omega} = \frac{1}{N}\sum_r c(r) \, .
\end{equation}
where the last sum is over the position of $N$ unit cells (with unit area) and $c(r)$ is the local Chern marker~\cite{bianco2011mapping,kitaev2006anyons,marrazzo2017locality,spataru2025topological}. We define the coarse-grained local Chern marker as,
\begin{equation}
    \tilde{c}_L(r) = \sum_\rho K_L(r-\rho) c(\rho) \, ,
\end{equation}
where the normalized coarse-graining function, $\sum_\rho K_L(\rho)=1$, is a gaussian with spread of $L$.
Coarse-graining introduces scales of observation: the coarse-grained local marker is blind to features smaller than the resolution $L$. We thus use the coarse-grained marker to assign a scale-dependent topological character to the HNL. We illustrate this in the example presented in Fig.~\ref {Fig:implement}J, which is a 3rd order nested $\text{AQH}^{\text{AQH}^\text{AQH}}$ lattice of size $7 \tx 7^{{4\tx 4}^{6\tx 6}}$, characterized through the lens of a coarse-grained local marker at zero energy. Color encodes the coarse-grained Chern marker at each point. We note that in a finite lattice, the total Chern number vanishes; the bulk and edge contributions to the Chern number carry opposite values. Accordingly, the hybrid bulk–edge structure is visible in this diagnostic. As the coarse-graining scale increases (from bottom to top in Fig.~\ref {Fig:implement}J), the bulk Chern number flips \(+1 \to -1 \to +1\). This behavior evidences a scale-dependent topological character. The bulk-edge configurations likewise evolve with scale. A study of the scale-dependent topological robustness in HNLs is presented in~\SMsec{sm:robust}.

By treating the different scales as distinct subspaces, one can return to Eq.~\eqref{2ndC} and associate a higher-order Chern number to the whole system, in accordance with the framework of the modern theory of polarization~\cite{benalcazar_quantized_2017}. For an \(n\)th-order HNL with scale-dependent Chern characters \(C_1^{C_2^{.^{.^{.}}}}\), the higher-order invariant reduces to
\begin{equation}
    C^{(n)} = \prod_i C_i \, .
\end{equation}
Consequently, if \(|C_i|>1\) at multiple scales, the nested system can host an amplified higher-order Chern number, \(|C^{(n)}|\gg 1\). We note that our approach contrasts with significant work in topological charge pumps where second/third Chern numbers in 4D/6D quantum Hall systems have been studied~\cite{petrides_six-dimensional_2018, zilberberg_photonic_2018, lohse_exploring_2018,chen2023second}. In such topological pumping schemes, the full high-dimensional dynamics are not captured since they represent slowly-varying snapshots of the Hamiltonian at specific quasimomenta $\vec k$ obtained by varying system parameters, and the entire lattice, in either momentum space or real space, is not realized simultaneously. Our hierarchically nested approach can instead realize the high-dimensional Hamiltonians, their corresponding higher-order Chern number introduced above, and their full dynamics on potentially 2D planar circuits of photonic ring resonator arrays.


\section{Magic Flat Bands} \label{sec:Magic}

Examining the energy-level distribution versus \(\alpha=J_2/J_1\) (Fig.~\ref{Fig:Band}A), we identify three regimes of band behavior as \(\alpha\) is tuned: (i) continuous deformation that largely preserves band shape; (ii) point-wise band touching that triggers a topological transition, e.g., a change \(|\Delta C|=1\); and (iii) bandwidth collapse (``spectral compression'') to zero, where levels condense to a point at a specific \(\alpha\) (a density-of-states (DOS) divergence) and then re-expand. These regimes recur across the \(\alpha\) axis. We refer to the \(\alpha\) at which the DOS diverges as the “magic” point, by analogy to magic-angle twisted bilayer graphene, where flat bands spanning the Brillouin zone appear only at specific twist angles. Many quantum systems routinely exhibit regime (i), with band shapes largely preserved under parameter tuning. An extreme case is provided by Landau levels—flat bands in electronic systems under magnetic fields~\cite{landau1930} (and in phononic/photonic analogs under strain~\cite{guglielmon2021landau,cheng2024three,barsukova2024direct})—which remain flat as, for example, the magnetic flux is varied. Topological insulators often display both regimes (i) and (ii). Regime (iii) is more singular and is famously realized in twisted bilayer graphene, where the tuning parameter is the twist angle. These flat bands innately differ from Landau levels because they only exist for a measure-zero set of the tuning parameters—the magic points—rather than continually.

In an infinitely large system (the thermodynamic limit), Landau levels are binary—they either exist or they do not. Their existence relies on the existence of a nonzero background field, not its size. Likewise, graph-theoretic flat bands—e.g., in Sutherland~\cite{sutherland1986localization}, Lieb~\cite{lieb1989two}, or line-graph lattices~\cite{kollar2020line}—passively persist so long as the underlying graph structure is preserved. Aharonov-Bohm cages similarly tie flatness to a fixed flux pattern; holding the flux constant preserves the flat bands. By contrast, the magic flat bands here share none of these traits.
This indicates that realizing such magic flat bands requires a nontrivial architectural complexity; importantly, they do \textit{not} arise in a single ring, a uniform ring lattice, or a simply connected block lattice.

A system that hosts these magic flat bands consequently has the following features:
\begin{itemize}
    \item Band flatness (similarly, curvature) is tunable by $\alpha$. In our $4\tx 4^{\infty\tx\infty}$ $\text{AQH}^\text{AQH}$ lattice, for example, near $\alpha \approx 0.87$, the band curvature varies slowly with $\alpha$, so the flat band does not rely on fine-tuning (Fig.~\ref{Fig:Band}B).
    \item The band curvature reverses sign across the flat band point.
    \item Some flat bands are bounded by trivial gaps (isolated perfect flat bands, Fig.~\ref{Fig:Band}E, third panel), others by non-trivial gaps (topologically embedded perfect flat bands, Fig.~\ref{Fig:Band}E, bottom panel). The latter is a rare emergence, e.g., found in topological HNLs.
    
    \item At the magic point, the flat band ensures the dominance of interactions. Photon-photon or photon-electron interactions can renormalize the band curvature; In some cases, this implies the existence of saddle points preferred by the energetics of the system, enabling ordered phases such as localizing phase transitions~\cite{parhizkar2024localizing}.
    \item The number of magic points grow with the $H_i$ lattice size.
    \item The spatial extent of the localized particles/wave-functions has a lower bound set by the size of $H_i$.
\end{itemize}
The magic behavior can be modeled by the coupled-mode formalism~\cite{hafezi2011robust}. As an example, we couple the edge modes of the $H_i$ to one another via $H_f$ (here, an AQH lattice). The coupled-mode schematic and its band structure are shown in~\SMsec{sm:analytic}; the resulting band structure is shown in Fig.~\ref{Fig:Band}. The calculated Chern numbers exactly mirror those of the hierarchical lattice. Each edge state is treated as a ring resonator mode that accumulates a half-round-trip phase of $e^{i\pi n}$, with $n$ the edge-mode index. The AQH-like couplings are then used to couple modes between different ``rings''. The effective coupling phases are modified by the half-round-trip phases. This is captured compactly by AQH Bloch Hamiltonians with momentum shifts of $0$, $\pi/2$, or $\pi$.


\section{Proposed circuits for photonic implementation in ring arrays}

We present scalable photonic circuits that implement QuMorph and HNLs in commercially available coupled ring-resonator arrays. We note that our choice of the resonator array platform is motivated by tremendous recent progress in foundry-compatible large-ring-array fabrication ~\cite{mittal2019photonic,flower2024observation}, control~\cite{zhang2019electronically,menssen2023scalable}, and programmability ~\cite{dai2024programmable,ma2024anisotropic,Mittal2016}, and spatial/spectral/dispersion optical spectroscopy techniques ~\cite{flower2024observation,xu2025chip,mehrabad2025multi} that suggest experimental realization of our proposed circuits and signatures of QuMorph is readily accessible. Fig.~\ref{Fig:implement}A–I summarizes the designs; for clarity, we consider only AQH-type HNLs (scaled up to third-order nesting). Design parameters, experimental considerations, and photonic Hamiltonians are provided in~\SMsec{sm:exp},~\SMsec{sm:analytic},~\SMsec{sm:scale}~\SMsec{sm:complete}, and~\SMsec{sm:latham}.

The circuits consist of ring arrays in an add-drop configuration, in which the main ``site’’ rings (black) are coupled by detuned ``link’’ rings that set the hopping rates (white, blue, and red for \(J_1\), \(J_2\), and \(J_3\), respectively). We start from a first-order \(4\tx 4\) nested AQH lattice (Fig.~\ref{Fig:implement}C). We then construct a \(4\tx 4^{4\tx 4}\) \(\text{AQH}^{\text{AQH}}\) HNL (Fig.~\ref{Fig:implement}E). Finally, we extend to a third-order HNL, constructing a \(4\tx 4^{4\tx 4^{4\tx 4}}\) \(\text{AQH}^{\text{AQH}^{\text{AQH}}}\) circuit (Fig.~\ref{Fig:implement}G).


\section{HNLs with hybrid topological phases}

\begin{figure*}[t]
    \centering
    \includegraphics[width=0.85
    \textwidth]{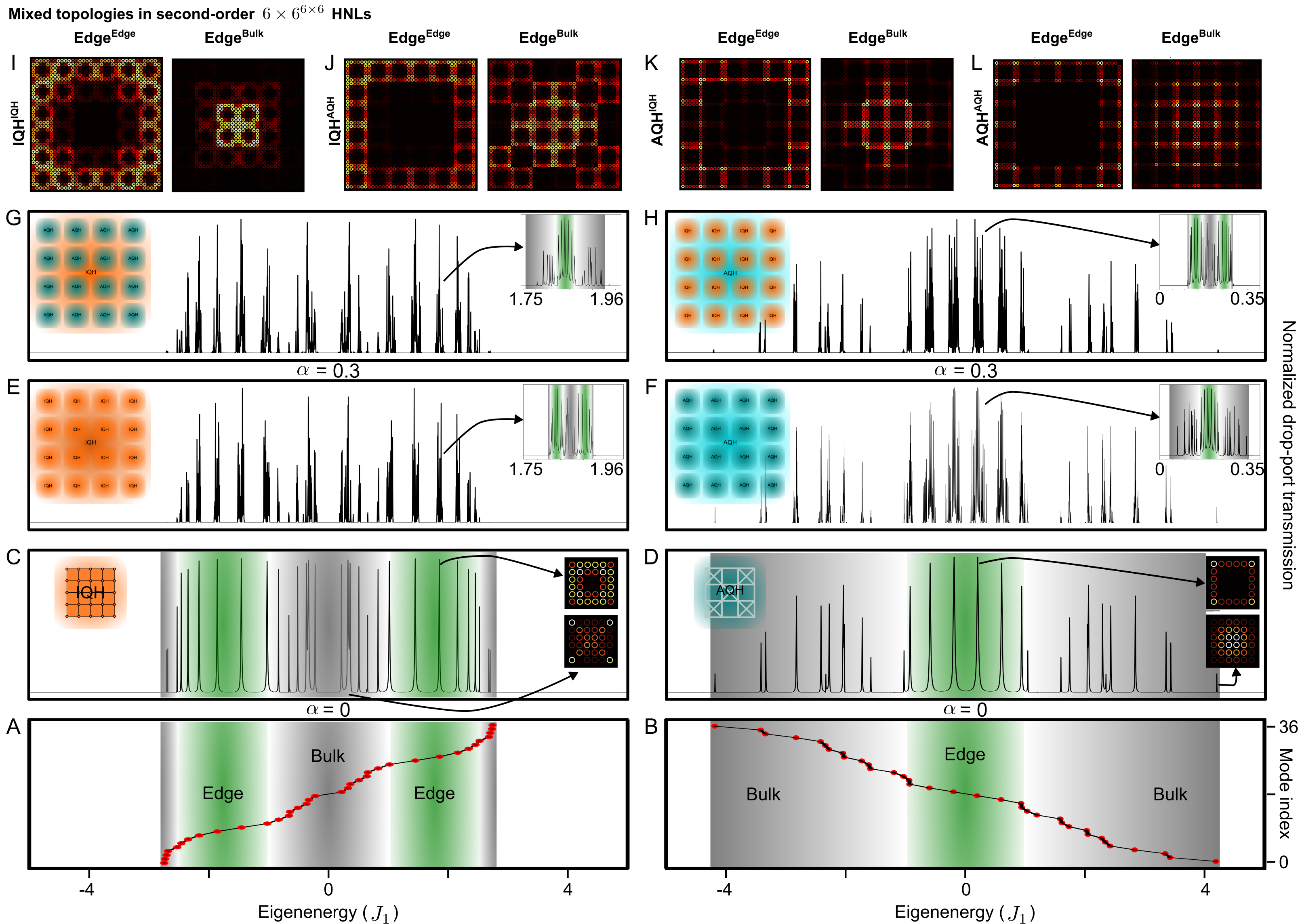}
    \caption{Calculated eigenvalues and drop-port transmission for \(6\tx 6\) photonic coupled-ring HNLs in an add–drop configuration: (A and C) IQH and (B and D) AQH. Insets show the indicated edge and bulk mode profiles. Drop-port transmission for \(6\tx 6^{6\tx 6}\) HNLs: (E) \(\text{IQH}^{\text{IQH}}\), (F) \(\text{AQH}^{\text{AQH}}\), (G) \(\text{IQH}^{\text{AQH}}\), and (H) \(\text{AQH}^{\text{IQH}}\). Insets mark bulk (gray) and edge (green) sections of the nested frequency bands. (I-L) Representative hybrid spatial-mode profiles, illustrating \(\text{edge}^{\text{edge}}\) and \(\text{edge}^{\text{bulk}}\) states in each HNL type. 
    }
    \label{Fig:mixed}
\end{figure*}

In the previous section, we focused on mixing two identical topological model (i.e., $\text{AQH}^\text{AQH}$). Next, we assess the capability of HNLs to host hybrid topological phases, by prototypically examining different AQH/IQH mixing configurations in a single HNL (Fig.~\ref{Fig:mixed}).
\begin{figure*}[t]
    \centering
    \includegraphics[width=0.85\textwidth]{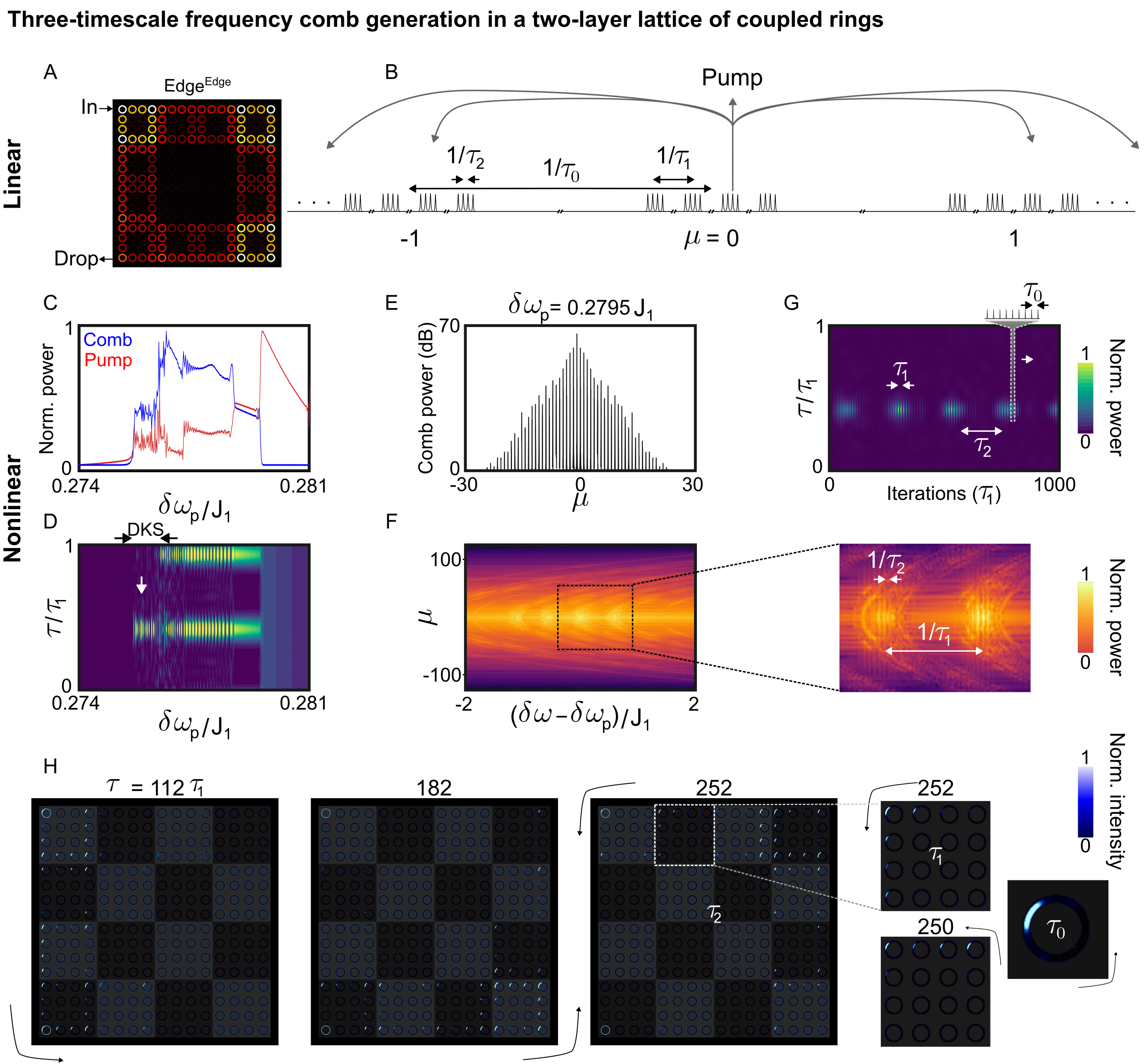}
    \caption{(A) Field intensity profile of a typical \(\text{edge}^{\text{edge}}\) mode in a \(4\tx 4^{4\tx 4}\) \(\text{AQH}^{\text{AQH}}\) HNL in an add–drop configuration. (B) Linear-regime drop-port transmission schematic as a function of single-ring longitudinal mode number $\mu$. Three mode spacings, corresponding to the three timescales of the \(\text{edge}^{\text{edge}}\) modes, are marked. Pumping one such mode can seed cascaded four-wave mixing (FWM) and comb generation (arrows). (C) Above–OPO-threshold pumping of the \(\text{edge}^{\text{edge}}\) mode marked in (B), versus pump frequency. Red (blue) indicates pump power (integrated comb power, excluding the light in the $\mu$ = 0 mode) in the HNL. (D) Corresponding spatiotemporal dynamics. The white arrow marks a single super-nested soliton at pump frequency $\delta\omega_{p} = 0.2795\,J_1$. (E) Comb profile of the soliton state. (F) Comb spectrum reorganized by single-ring mode spacing and first-order nested frequency. Inset: three-timescale mode locking. (G) Temporal output showing pulse bursts with three periodicities. (H) Soliton intensity distribution and counter-clockwise (CCW) circulation around the array. The three phase-locked timescales are indicated. For clarity, only site rings are shown.  
    }
    \label{Fig:Comb}
\end{figure*}
We begin with two \(6\tx 6\) first-order nested lattices, one IQH-like and one AQH-like. Fig.~\ref{Fig:mixed}A–D shows eigenvalues and drop-port transmission, highlighting bulk (gray) and edge (green) sectors and representative field profiles for this lattice. We then construct the four possible IQH/AQH mixtures in a \(6\tx 6^{6\tx 6}\) HNL and present the corresponding drop-port spectra and hybrid \(\text{edge}^{\text{edge}}\)/\(\text{edge}^{\text{bulk}}\) profiles (Fig.~\ref{Fig:mixed}E–L). The characteristic features of the constituent phases persist in a nesting-order–dependent fashion: IQH shows two edge bands separated by a bulk band, whereas AQH shows two bulk bands separated by an edge band; AQH profiles are more edge-confined due to its large gap size (Fig.~\ref{Fig:mixed}E–H, insets). These observations demonstrate controllable spatial profiles, spectra, and dispersion in HNLs with lattice-order–dependent synthetic gauge fields. Importantly, the distinct spatial and spectral signatures in Fig.~\ref{Fig:mixed} provide clear observables for experimental validation of QuMorph in HNLs (\SMsec{sm:eigenmixed}).

Beyond AQH/IQH mixing, many other ring-resonator-compatible quantum Hall phases are accessible on this platform (\SMsec{sm:othertopo}). For example, spin and valley Hall phases can be realized via unit-cell symmetry breaking in honeycomb lattices~\cite{yang2021optically}. Higher-order topological phases can likewise be implemented in coupled ring resonators~\cite{mittal2019photonic}. Recently introduced ring-based Floquet~\cite{hashemi2024floquet}, non-Hermitian~\cite{hashemi2025reconfigurable}, and hyperbolic~\cite{huang2024hyperbolic} topological lattices are also available. This suggests that mixing—and probing the interplay—of more than two phases is feasible in third-order HNLs.


\section{Beyond linear regime: QuMorph for integrated nonlinear optics}

Beyond the linear regime, QuMorph in HNL can provide a rich new platform to explore multi-scale spatio-temporal nonlinear phenomena, of which, in this section, we show an example for integrated nonlinear optics. We note that the diverse range of hybrid new states shown in Fig.~\ref{Fig:implement} exhibits a variety of spatial profiles that may support complex nonlinear mode-locked states.
This spatial degree of freedom provides a connection to spatial cavity solitons, which have been studied for three decades~\cite{LugiatoPhys.Rev.Lett.1987} as solutions for an all-optical memory~\cite{CoulletChaosJournalofNonlinearScience2004} and optical computing~\cite{PedaciAppliedPhysicsLetters2006} using addressable light spots on two-dimensional planes~\cite{BarlandNature2002}.
However, experimental sensitivity to fabrication imperfections has challenged widespread deployment, spurring research into temporal cavity solitons.
Our HNL architecture combines both temporal and spatial concepts through multi-temporal mode locking and addressable, reconfigurable localization of DKS pathways with intrinsic topological robustness, which may mitigate limitations of conventional cavity solitons. Moreover, our multi-timescale system simultaneously offers a large spectral bandwidth from the small microring (large FSR) housed in a large super-mode volume.
Therefore, our system spans a large spectral window for potential self-referencing use~\cite{ApolonskiPhys.Rev.Lett.2000,JonesScience2000} while providing smaller thermo-refractive bistability and noise~\cite{HuangPhys.Rev.A2019} than single microring operation~\cite{MoilleOptica2025}, potentially enabling metrological applications.

We illustrate an example of such exotic states by studying QuMorph multi-timescale frequency-comb generation, which is of continuous broad interest in theory~\cite{mittal2021topological,tusnin2023nonlinear,hashemi2024floquet,hashemi2025reconfigurable,huang2024hyperbolic} and experiment~\cite{flower2024observation,xu2025chip,mehrabad2025multi}. Specifically, we go beyond prior nested comb studies to explore multi-timescale, mode-locked solitonic solutions with exotic dynamics. Fig.~\ref{Fig:Comb} summarizes the results. The HNL is \(4\tx 4^{4\tx 4}\) \(\text{AQH}^{\text{AQH}}\) with \(\alpha=0.3\). Representative linear-regime \(\text{edge}^{\text{edge}}\) field profiles and drop-port transmission are shown in Fig.~\ref{Fig:Comb}A–B, revealing three characteristic mode spacings that reflect the HNL’s multi-timescales (single-ring free spectral range (FSR), first-order nesting FSR, and second-order nesting FSR, corresponding to $1/\tau_0$, $1/\tau_1$, and $1/\tau_2$, respectively. Here, for \(\alpha=0.3\), the ratio of these timescales in the simulations is approximately $\tau_2:\tau_1:\tau_0=20000:100:1$).
 
For comb-generation simulations, we use the Lugiato–Lefever equation (LLE) formalism~\cite{mittal2021topological} (\SMsec{sm:nonl}). Similar to single-resonator dynamics, a sufficiently powerful pump laser triggers degenerate nonlinear four-wave mixing in the HNL, leading to optical parametric oscillation (OPO) when properly phase-matched (\textit{i.e.,} anomalous dispersion~\cite{RazzariNaturePhoton2010} or complex normal dispersion with high-order dispersion phase-matching~\cite{LuOptica2019} or complex photonics engineering~\cite{MoilleCommunPhys2023}).
This process cascades to form integrated frequency combs and, under proper conditions, can enter a mode-locked regime such as that of dissipative Kerr solitons (DKSs)~\cite{KippenbergScience2018}.
Moreover, topological ring arrays enable highly relaxed dispersion conditions for phase matching~\cite{mehrabad2025multi}, so DKS formation from soft-excitation~\cite{MatskoPhys.Rev.A2012} nonlinearity of an original OPO may rely on different dispersion conditions---thanks to the resonance-nesting of our system (Fig.~\ref{Fig:Comb}B)---than single-resonator systems.
Importantly, such novel and exciting nonlinear solutions in our HNL system are accessible with conventional adiabatic pathways simply by sweeping the pump frequency across a resonance (Fig.~\ref{Fig:Comb}C). The single-resonator is considered to be in the anomalous dispersion regime; therefore, multi-timescale mode-locked states may exist and may be accessible by selecting the pump frequency accordingly (Fig.~\ref{Fig:Comb}C–D).

Spectral and temporal characteristics of a representative DKS are shown in Fig.~\ref{Fig:Comb}E–G, with clear signatures of three-timescale mode locking. Fig.~\ref{Fig:Comb}H shows temporal snapshots of the DKS, revealing the spatial intensity distribution of the three-timescale pulse (Supplementary Movie 6). Each timescale is independently tunable via the scale-dependent hopping rates (i.e., via single FSR, $J_1$, and $J_2$, respectively).

Taken together, the observed comb spectra with scale-tunable nested tooth families, stable pulse trains with nested envelopes, and \(\text{edge}^{\text{edge}}\)-localized profiles that trace the lattice hierarchy establish QuMorph in HNLs as a scalable route to multiscale mode locking and programmable soliton synthesis for on-chip nonlinear photonics.



\section{Outlook}

On a fundamental level, QuMorph and HNLs can elevate multi-scale synergy to a programmable control knob, defining continuous “metamorphosis paths’’ between nested-lattice Hamiltonians. In contrast to approaches based on moir\'{e} structures (where emergent phenomena are controlled by the superlattice period), here the structure geometry can be kept fixed, with a readily tunable coupling parameter determining the spectral properties. Looking ahead, several intriguing directions stand out: (i) classifying such QuMorph paths with scale-dependent markers---coarse-grained local Chern indicators and a hierarchical counterpart of conventional Chern vectors~\cite{liu2022topological}---that predict when gap inversions force isolated edge manifolds; (ii) sharpening a criterion for magic flat bands at the cocoon stage into quantitative predictions for the magic values of \(\alpha\) and their scaling with nesting order; (iii) extending QuMorph to driven, non-Hermitian, and weakly interacting regimes to test whether cocoon mini-gaps seed robust lasing, synchronization, or interaction-stabilized phases. Another intriguing avenue is the exploration of HNLs as a platform for exploring self-similarity in the ultra-fast nonlinear optics regime, such as similaritons~\cite{dudley2007self}.

On the applications and devices side, realization of our proposed foundry-compatible HNLs in coupled ring resonators provides an immediate platform (we note also the highly programmable optomechanical networks~\cite{slim2025programmable} as another option): \(\alpha=J_2/J_1\) can be tuned thermo/electro-optically, and transmission/group-delay/spatial spectroscopy maps QuMorph directly. Towards this direction, three tracks stand out: programmable dispersion and routing (scale-addressable edge-bulk solutions for filters, delay lines, robust interconnects); interaction enhancement and nonlinear control (magic flatbands~\cite{martinez2023flat} and hybrid confinement for multiscale combs~\cite{flower2024observation}, hyper-nested mode-locking~\cite{xu2025chip}, ultra-relaxed frequency-phase matching~\cite{mehrabad2025multi}, edge-of-edge solitons); and dual-comb nested spectroscopy and sensing/signal processing (detached edge bands as high-\(Q\), low-background channels with localized responsivity). QuMorph thus offers a concise blueprint for co-designing topology, dispersion, and nonlinearity across nested orders with a single, scalable knob.

\section{Acknowledgements}

The authors wish to acknowledge fruitful discussions with Yanne Chembo, Sachin Vaidya, Supratik Sarkar, Apurva Padhye, Chao Li, and Sunil Mittal. 


\section{Competing interests}
M.J.M., A.P., L.X., and M.H. have filed a provisional patent covering QuMorph processes and HNL architectures. The authors declare no other competing interests.

\section{Data and materials availability}
All of the data that support the findings of this study are reported in the main text and Supplementary Materials. Source data are available from the corresponding authors on reasonable request.

\newpage
\section*{Supplementary Materials for\\ Quantum metamorphosis: Programmable emergence and the breakdown of bulk-edge dichotomy in multiscale systems}


\section*{Supplementary Table of Contents}
\vspace{-0.75em}
\hrule
\vspace{0.9em}

\newcommand{\SupTOCEntry}[3]{%
  #1.\quad\hyperref[#2]{#3}%
  \dotfill\pageref{#2}\par
}

\begingroup
\setlength{\parindent}{0pt}
\setlength{\parskip}{0.35em}


\SupTOCEntry{S1}{sm:exp}
  {Considerations for experimental realization of QuMorph in HNLs with CMON chips}
\SupTOCEntry{S2}{sm:stat}
  {Example for Particle Statistics}
\SupTOCEntry{S3}{sm:bandev}
  {Band evolution}
\SupTOCEntry{S4}{sm:qumorphall}
  {QuMorph in different mixing configurations}
\SupTOCEntry{S5}{sm:flatband}
  {Flat band criterion}
\SupTOCEntry{S6}{sm:analytic}
  {Analytical description of QuMorph in HNLs}
\SupTOCEntry{S7}{sm:othertopo}
  {First-order nested topological models and lattices studied with coupled rings}
\SupTOCEntry{S8}{sm:robust}
  {Robustness}
\SupTOCEntry{S9}{sm:scale}
  {Scale up}
\SupTOCEntry{S10}{sm:eigen}
  {Eigenvalues versus drop-port spectra in AQH--AQH HNLs}
\SupTOCEntry{S11}{sm:eigenmixed}
  {Eigenvalues versus drop-port spectra in mixed-topology HNLs}
\SupTOCEntry{S12}{sm:complete}
  {Complete sets of field profiles}
\SupTOCEntry{S13}{sm:large}
  {QuMorph at large $\alpha$}
\SupTOCEntry{S14}{sm:fract}
  {Tunable Fractal Dimension}
\SupTOCEntry{S15}{sm:latham}
  {Lattice Hamiltonian}
\SupTOCEntry{S16}{sm:nonl}
  {Nonlinear simulations in HNLs}



\SupTOCEntry{Supplementary Movie 1}{sm:nonl}
  {AQH-AQH QuMorph}
\SupTOCEntry{Supplementary Movie 2}{sm:nonl}
  {AQH-IQH QuMorph}
\SupTOCEntry{Supplementary Movie 3}{sm:nonl}
  {IQH-IQH QuMorph}
\SupTOCEntry{Supplementary Movie 4}{sm:nonl}
  {QuMorph band evolution}
\SupTOCEntry{Supplementary Movie 5}{sm:nonl}
  {QuMorph wavefunction profiles}
\SupTOCEntry{Supplementary Movie 6}{sm:nonl}
  {Three-timescale spatio-temporal soliton evolution}

\endgroup



\begin{supplement}

\newpage


\section{Considerations for experimental realization of QuMorph in HNLs (CMOS-compatible ring–resonator arrays)}\label{sm:exp}

We outline notes and considerations for the experimental realization of QuMorph in HNLs. We target commercially available platforms (silicon (Si), silicon nitride (SiN), and lithium niobate (LiN)), which enable volume-manufacturability at a wafer scale. Moreover, we focus on central operational bandwidths at telecommunications wavelengths due to the existence of mature optical spectroscopic techniques.
We note that our choice is in part motivated by recent progress on wafer-scale functional device yield of large first-order nested ring arrays of up to $10\times10$ coupled ring lattices, as well as established transport, dispersion, and spatial imaging techniques both in the linear and nonlinear regimes~\cite{flower2024observation,xu2025chip,mehrabad2025multi}. Moreover, we also note that beyond these platforms, multiple alternative choices are also feasible, including large ring arrays in AlGaAs~\cite{pang2025versatile}. Beyond such photonic systems, opto-mechanical networks of large lattice size have been recently realized and can be an immediate alternative platform~\cite {slim2025programmable}.

\subsection*{Platform}

For the fabrication of photonics ring resonators in platforms such as Si, SiN, and LiN, operation at telecommunication wavelength, a typical process stack uses 300~nm to 900~nm core thickness, and waveguide widths 900~nm to 1500~nm. We adopt TE polarization throughout. The design aims for low propagation loss ($<0.3$~dB/cm), critically coupled I/O rings, and $Q_{\rm int}$ in the few-million range.

\subsection*{Design}
The HNLs can be designed for foundry fabrication using conventional design approaches for large photonic ring arrays~\cite{flower2024observation,xu2025chip,mehrabad2025multi}. Moreover, very recently, artificial intelligence (AI) agentic frameworks have been introduced for photonic chip designs with large complexities exceeding hundreds of components~\cite{sharma2025ai}, which is comparable to the number of rings in HNLs. Therefore, such AI agents may enable even more efficient and versatile design capabilities for HNLs at a large scale. At the same time, the diversity of functionalities and broad parameter space for chip design for HNLs provides an intriguing testbed to benchmark the capabilities of such agentic AI frameworks for smart photonic devices.

\subsection*{HNL layout: $4\times4^{\,4\times4}$}
A second-order HNL places a $4\times4$ ``inner'' lattice at each site of a $4\times4$ ``outer'' lattice. Inner-lattice couplings define $J_1$; corner-to-corner interconnects between neighboring inner lattices realize the outer network with coupling $J_2$. The control parameter is
\[
\alpha \equiv \frac{J_2}{J_1},
\]
tuned thermo/electro-optically (via link-ring detuning and/or coupling gaps) to drive QuMorph.

\paragraph*{Footprint and pitch.}
Single-ring radius $R=20$~\textmu m to $30$~\textmu m yields a single-ring FSR
\[
\Omega_F/2\pi \approx \frac{c}{2\pi n_g L} \approx (500\text{ to }1000)~\mathrm{GHz}
\]
(for $n_g\approx1.9$ and $L\!=\!2\pi R$). A practical inner-lattice pitch $p_{\rm in}=$ (40 to 60)~\text{\textmu m} sets a $4\times4$ tile size of $\approx(160\text{ to }240)~\text{\textmu m}$ per side. The outer-lattice pitch $p_{\rm out}$ equals the inner-tile diagonal plus routing margin [$p_{\rm out}\!=\! \sqrt{2}\,4p_{\rm in} + (40$ to $80$)~\textmu m], keeping the full $4\times4^{\,4\times4}$ HNL within a few~mm$^2$.

\subsection*{Implementing IQH/AQH building blocks and \texorpdfstring{$\alpha$}{alpha} control}
\paragraph*{Synthetic gauge (AQH) with detuned link rings.}
Nearest-neighbor site–site coupling is mediated by off-resonant ``link'' rings (path-length offset $\Delta L$) that imprint direction-dependent phases (e.g., $\pm\pi/4$) and set the hopping amplitude. Next-nearest neighbor couplings are realized with zero-phase links. This yields the anomalous quantum Hall (AQH) lattice used in the main text. Integer-QH-like variants (uniform synthetic flux per plaquette) can be implemented with suitable phase layouts.

\paragraph*{Setting $J_1$ and $J_2$.}
For both inner and outer networks, the effective coupling $J$ is controlled by the bus–link and link–site directional-coupler gaps ($g$) and the link-ring detuning $\delta\omega_{\rm link}$:
\[
J \simeq \frac{\kappa_{\rm eff}}{2}\,\frac{\Delta}{\Delta^2+(\delta\omega_{\rm link}/2)^2},
\]
where $\kappa_{\rm eff}$ is the effective waveguide–ring coupling rate and $\Delta$ encodes the designed coupling phase path. Practically, fabricate $J_1$ as fixed (inner tiles) and make $J_2$ tunable via integrated heaters on the outer-lattice link rings (or phase shifters on corner waveguides), providing $\alpha$ control over at least an order-of-magnitude range (e.g., $\alpha\in 0.1\!-\!3$).

\subsection*{Representative geometry and targets}
\begin{center}
\begin{tabular}{l l}
\toprule
Parameter & Target range (telecom band)\\
\midrule
Waveguide width $\times$ height & ($1.1$ to $1.3)~\text{\textmu m} \times (0.75$ to $0.85)~\text{\textmu m}$ \\
Ring radius $R$ & ($20$ to $30$)~\text{\textmu m}$\;\;(\Omega_F/2\pi\approx(0.5$ to $1$)~THz) \\
Inner coupling $J_1/2\pi$ & ($1$ to $5$)~GHz \\
Outer coupling $J_2/2\pi$ & ($0.3$ to $8$)~GHz (tunable) \\
Ext. coupling per I/O $\kappa_{\rm ex}/2\pi$ & $0.05$–$0.3\,J_1$ (add–drop) \\
Intrinsic loss $\kappa_0/2\pi$ & $<50$~MHz (SiN, low-loss stack) \\
Link-ring detuning & ($5$ to $30$)~GHz off site-ring line \\
Heater shift (per $\pi$ phase) & $<20$~mW per link (thermo-optic) \\
\bottomrule
\end{tabular}
\end{center}

\subsection*{Fabrication notes (wafer-scale)}
\begin{itemize}
  \item \textbf{Coupler gaps.} Using a small set of discrete gaps (e.g., 250~nm, 300~nm, 350~nm, 400~nm) can bracket $J_1$ and $J_2$; particularly, including two-ring and four-ring test structures to calibrate $J(g)$ post-fab is useful.
  \item \textbf{Cladding.} Glass cladding helps suppress parasitic coupling and avoid dirt, and maintains coupler gaps across the array.
\end{itemize}

\subsection*{Linear characterization: pump, probe, and transport}
\paragraph*{Add–drop mapping.}
A narrow–linewidth tunable laser can be used to sweep the telecom band. Recording input, through, and drop ports enables extraction of $J_1$, $J_2$, $\kappa_{\rm ex}$, $\kappa_0$ by fitting to theory. Mapping the edge band can be performed by injecting at a boundary site (corner ring) using an input bus waveguide and scanning the pump frequency.

\paragraph*{Group-delay and band tomography.}
Using an off-the-shelf optical vector analyzer, group-delay $\tau_g = \partial\phi/\partial\omega$ can be directly measured from the add-drop ports of the HNL devices. By plotting $(\omega,\tau_g)$, one can visualize band edges and gap openings as $\alpha$ is tuned. Stepping $\alpha$ via heater bias on outer links can reveal the \emph{butterfly} $\to$ \emph{cocoon} $\to$ \emph{mantis} crossover.

\paragraph*{Edge vs bulk transport.}
Launching from a corner-ring input port (edge-preferential excitation) can contrast the spatial properties of the modes. One can quantify edge localization length vs.\ frequency and $\alpha$; identify isolated edge bands by the coexistence of strong edge transmission and suppressed bulk throughput inside trivial gaps.

\subsection*{Real-space imaging}
Top-view imaging of the out-of-plane scattered with a typical $20\!\times$ to $50\!\times$ objective can directly reveal the spatial signatures of modes.

\begin{itemize}
  \item \textbf{Edge maps.} At frequencies inside topological gaps, one should observe unidirectional edge confinement; in the cocoon, hybrid profiles should be expected (\(\text{edge}^{\text{bulk}}\), \(\text{bulk}^{\text{edge}}\)).
  \item \textbf{Isolated edge bands.} Inside ``residue-topology'' windows, one should look for detached edge manifolds (edge-only loops) with minimal bulk background.
  \item \textbf{Magic flat bands.} Near flat-band points, the bulk becomes non-propagating; imaging shows stationary bulk intensity with dispersive edge channels at the same frequency.
\end{itemize}

\subsection*{Disorder budget and robustness}
A key consideration for the realization of topological ring arrays is the degree of robustness against disorder. Let $U$ denote a typical ring-to-ring resonance inhomogeneity (from width/thickness variations and disorder). It is crucial to maintain $U \lesssim 2J_1$ for robust edge transport; if wafer disorder is larger, increasing $J_1$ (smaller coupler gaps) and re-centering $J_2$ by heater tuning to preserve the target $\alpha$ can maintain the device functionality. Include perimeter guard rings and dummy fill to stabilize etch proximity effects.

\subsection*{Excitation and detection operating modes and recommended flow}
\begin{enumerate}
  \item \textbf{Post-fab calibration:} Single- and two-ring test devices can be used to extract group index $n_g(\lambda)$, loss rate $\kappa$, and hopping rate $J$, to inform the array design.
  \item \textbf{First-order HNL:} individual $4\times4$ devices can be probed to confirm their band edges and gap sizes, to confirm $J_1$ targets.
  \item \textbf{Set $\alpha$:} (optional) Implementing and biasing 2nd-order HNL link-ring heaters can control dial $J_2$; sweeping $\alpha$ in coarse steps to locate the cocoon regime, then fine-stepping to resolve mini-gaps and flat-band points.
  \item \textbf{Spectral–spatial mapping:} For each $\alpha$, add–drop spectra, group-delay, and real-space images can be recorded and assembled to show spatio-spectral signatures of QuMorph discussed in the main text.
\end{enumerate}

\subsection*{Notes on nonlinear optical experiments in HNLs}
For Kerr-nonlinear tests (comb/soliton regimes), the HNL can be excited in the quasi-CW pumping regime~\cite{flower2024observation} near an edge resonance. Broadband OSAs covering the telecommunication band can reveal the comb spectra. IR cameras can be used to directly image the generated comb light by filtering the pump prior to the camera. Nested timescales can be directly measured using heterodyne-based off-the-shelf OSAs, which provide ultra-high spectral resolutions of up to 40~fm~\cite{flower2024observation}.

\subsection*{Additional useful practical checklist}
\begin{itemize}
  \item Verification of link-ring detuning and phase imprint (nearest neighbors $\pm\pi/4$) across all tiles.
  \item Confirmation of $\alpha$-tuning range ($\geq 10\times$ dynamic range in $J_2$) with heater sweeps.
  \item Edge–bulk drop-port transmission intensity contrast inside topological gaps for multiple $\alpha$ values.
  \item Identifying isolated edge band and magic flat-band points via spectral + imaging diagnostics.
\end{itemize}


\section{Example for Particle Statistics} \label{sm:stat}

Given the eigenfunctions,
\begin{equation}
    \ket{\psi^{ab}(\vec{k})} = \ket{\psi^a_i(\vec{k}_i)}\ket{\psi^b_f(\vec{k}_f)} \, , \quad H \ket{\psi^{ab}(\vec{k})} = E_{ab}\ket{\psi^{ab}(\vec{k})} \equiv (E_a + E_b)\ket{\psi^a_i(\vec{k}_i)}\ket{\psi^b_f(\vec{k}_f)}\, ,
\end{equation}
define the annihilation operator below,
\begin{equation}
    c_{ab,\vec{k}} \equiv \sum_{\vec{r}_i,\vec{r}_f} \left[\psi^a_i(\vec{r}_i;\vec{k}_i)\psi^b_f(\vec{r}_f;\vec{k}_f)\right]^* c_{\vec{r}_i,\vec{r}_f} \, ,
\end{equation}
where $c_{\vec{r}_i,\vec{r}_f}$ annihilates a particle at $\ket{\vec{r}_i,\vec{r}_f}$ with $\vec{r}_{i,f}$ pointing at a position within $H_{i,f}$. Using the orthonormality of $\psi^{a,b}_{i,f}$ it is easy to check that for a fermion,
\begin{equation}
    \left\{c_{ab,\vec{k}},c^{\dagger}_{a'b',\vec{k'}} \right\} = \delta_{aa'}\delta_{bb'}\delta_{\vec{k}\vec{k}'} \, , \quad \text{and} \quad \left\{c_{ab,\vec{k}},c_{a'b',\vec{k'}} \right\}=0 \, ,
\end{equation}
and for a boson,
\begin{equation}
    \left[c_{ab,\vec{k}},c^{\dagger}_{a'b',\vec{k'}} \right] = \delta_{aa'}\delta_{bb'}\delta_{\vec{k}\vec{k}'} \, , \quad \text{and} \quad \left[c_{ab,\vec{k}},c_{a'b',\vec{k'}} \right]=0 \, .
\end{equation}
The grand canonical distribution functions for fermions ($+$) and bosons ($-$) are then given by,
\begin{equation}
    n^\pm_{ab}(\vec{k}) = \left\langle c^\dagger_{ab,\vec{k}} c_{ab,\vec{k}}  \right\rangle  = \frac{1}{e^{\beta(E_{ab}(\vec{k})-\mu)} \pm 1} \, ,
\end{equation}
with $\mu$ being the chemical potential and $E_{ab}$ being the energy of the eigenstates $\ket{\psi^a_i(\vec{k}_i)}\ket{\psi^b_f(\vec{k}_f)}$. Since tuning $\alpha$ changes $E_{ab}$, the distribution function is also going to depend on $\alpha$.

\newpage
\section{Band evolution}\label{sm:bandev}
Fig.~\ref{Fig:band} illustrates the calculated band structure evolution for a semi-infinite (finite in x, periodic in y) $\text{AQH}^\text{AQH}$ HNL as a function of $\alpha$. The first-order lattice is $4\tx 4$. As $\alpha$ is tuned from $0$ to larger values, the degeneracy of the bands of the HNL is lifted, with the bands gradually evolving and hybridizing as $\alpha$ increases, giving rise to a rich landscape of programmable band structure. Particularly, the emergence and evolution of perfect magic and topologically embedded flat bands and isolated edge bands can be clearly tracked. At very large values of $\alpha$, the band structure converges to a set of degenerate energy bands. 
\begin{figure*}[t]
    \centering
    \includegraphics[width=0.92
    \textwidth]{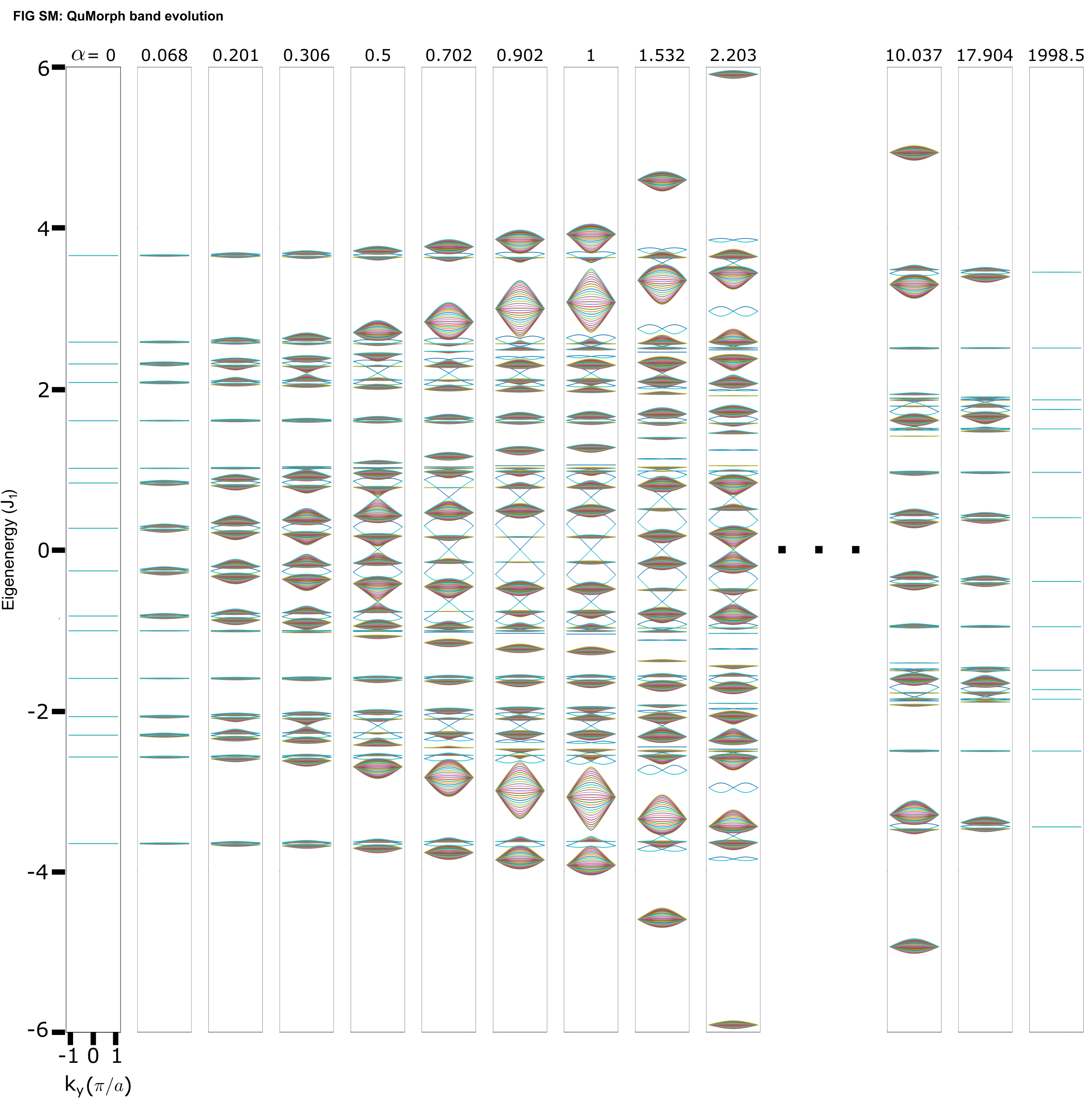}
    \caption{Band structure evolution during QuMorph in a semi-infinite $\text{AQH}^\text{AQH}$ HNL of size $4 \tx 4^{30 \tx \infty}$. As the tuning varies from $\alpha = 0$ to $\alpha \gg 1$ the spectrum evolves from isolated AQH orbitals to the corresponding metamorphosed one. In the cocoon stage, around $\alpha = 1$, orbitals mix and exotic single particle behaviors emerge.
    }
    \label{Fig:band}
\end{figure*}
%

\newpage
\section{QuMorph in different mixing configurations}\label{sm:qumorphall}

Fig.~\ref{Fig:fullmorph} illustrates the calculated QuMorph for three different mixing configurations between AQH and IQH phases in large finite-size 2nd-order HNLs.
\begin{figure*}[t]
    \centering
    \includegraphics[width=0.92
    \textwidth]{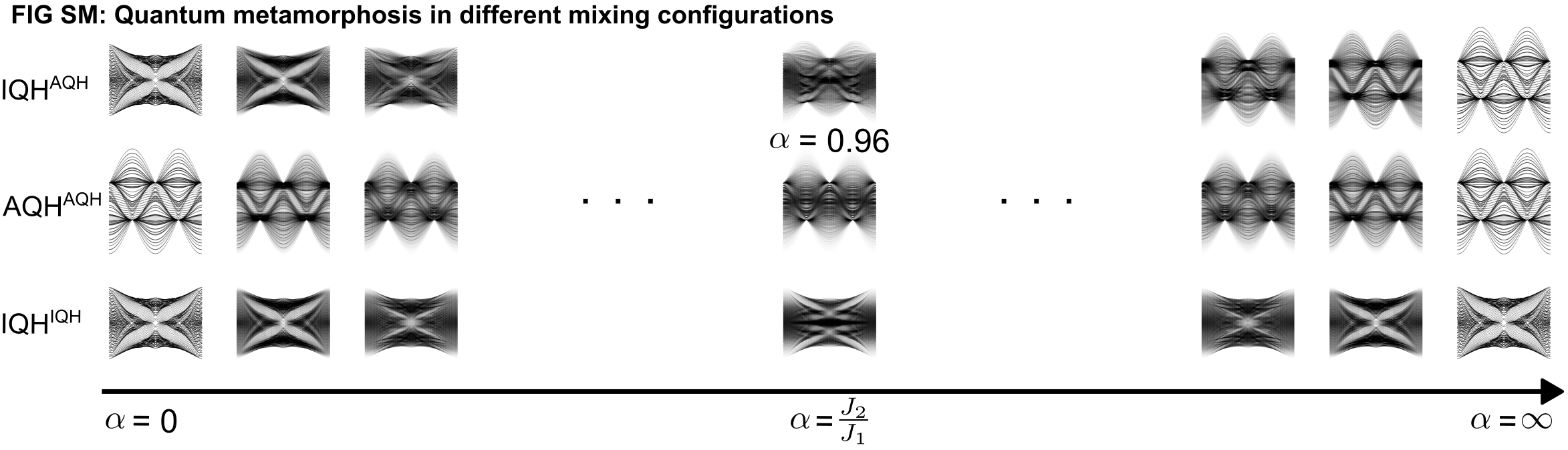}
    \caption{From top to bottom rows: QuMorph in large finite-size HNLs with $\text{IQH}^\text{AQH},$ $\text{AQH}^\text{AQH}$, and $\text{IQH}^\text{IQH}$ mixed-topological systems.
    }
    \label{Fig:fullmorph}
\end{figure*}

\newpage
\section{Flat band criterion}\label{sm:flatband}

The anomaly shows itself as the Jacobian of the chiral transformation,
\begin{equation}
    I = \int \mathcal{D}\bar{\psi}\psi e^{iS} \rightarrow \int \mathcal{D}\bar{\psi'}\psi' e^{i\mathcal{A}_5} e^{iS'} \, ,
\end{equation}
where $S$ is the $(4+0)$-dimensional action while $\bar{\psi}$ and $\psi$ are particle fields and $\mathcal{A}_5$ is the chiral anomaly or the logarithm of the Jacobian. Through a total chiral rotation $\psi$ goes back to itself, and thus we should have $\psi=\psi'$. In that case, we can write the above as,
\begin{equation}
    I =  I e^{i\mathcal{A}^T_5}  \, ,
\end{equation}
With $\mathcal{A}^T_5$ being the anomaly of the total chiral rotation. Therefore, the path integral $I$, or the flat band theory, is only realizable if,
\begin{equation}
    e^{i\mathcal{A}^T_5} \in 2\pi \mathbb{Z} \, .
\end{equation}
This is a criterion for forming flat bands in the system. Similar to the Chern number, the character of this criterion depends on $\alpha$. At the cocoon stage, for $H_{i,f}$ describing gauge theories (e.g., the Hall effect for massless particles), the total anomaly $\mathcal{A}_5^T$ is given by,
\begin{equation}
    \mathcal{A}_5^T = \frac{2\pi}{16\pi^2}\int d^2x_i d^2x_f \, \epsilon^{\mu\nu\rho\sigma}F^a_{\mu\nu}F^a_{\rho\sigma} \, ,
\end{equation}
where $F^a_{\mu\nu}$ is the non-Abelian field strength (which reduces to Abelian field strength for Abelian systems) and the integral is over the unit cells (i.e., the magnetic unit cells in the quantum Hall system) of the $H_{i,f}$ systems. Thus, we have a realization of a four-dimension chiral anomaly~\cite{fujikawa2004path,rylands2021chiral,parhizkar2024path,parhizkar2024generic,parhizkar2025field} and the fifth-dimensional parity anomaly.
Our simple example here, therefore, shows that increasing orders of metamorphosis and nesting manifest increasing Chern numbers (1st, 2nd, etc) and higher-dimensional anomalies.

\newpage
\section{Analytical description of QuMorph in HNLs}\label{sm:analytic}

Fig.~\ref{Fig:theory} demonstrates a theoretical model for how band crossing and perfect flat bands emerge for QuMorph in HNLs using a four-band nested Hamiltonian formalism. This theoretically calculated band structure evolution is in close agreement with the numerically calculated counterpart presented in Fig. 2 in the main text. 

The standard square lattice AQH Hamiltonian is given by~\cite{PhysRevLett.121.023901},
\begin{equation}
\label{eq:AQH}
  H_{\text{AQH}}\left(\vec{k}\right)=\left(
  \begin{array}{cccc}
   d_0\left(k_x,k_y\right) 
   + d_z\left(k_x,k_y\right) & d_x\left(k_x,k_y\right)-i d_y\left(k_x,k_y\right) \\
   d_x\left(k_x,k_y\right)+i d_y\left(k_x,k_y\right) & d_0\left(k_x,k_y\right) -d_z\left(k_x,k_y\right) 
  \end{array}
  \right) = \mathbf{1} d_0\left(\vec{k}\right) +  \vec{\sigma} \cdot \vec{d}\left(\vec{k}\right)  \, ,
\end{equation}
with the $d_{0,x,y,z}(k_x,k_y)$ defined as
\begin{align}
    d_0(k_x,k_y) &= \cos{k_x} + \cos{k_y} \, ,\\
    d_x(k_x,k_y) &= 2\sqrt{2}\cos{\frac{k_x}{2}}\cos{\frac{k_y}{2}} \, ,\\
    d_y(k_x,k_y) &= -2\sqrt{2}\sin{\frac{k_x}{2}}\sin{\frac{k_y}{2}} \, ,\\
    d_z(k_x,k_y) &= \cos{k_y}-\cos{k_x} \, ,
\end{align}
where we have normalized the lattice vector to $1$. 
In order to model the HNL, we aim to substitute each site with another AQH lattice. However, here we wish to see the emergence of topological gaps and flat bands, and for that, substituting each site with only the $\text{edge}^{\text{edge/bulk}}$ subspace of the AQH system suffices. We then treat the edge state as a field in a ring resonator with detuning $\Delta$ away from a resonance frequency. Distinguishing the two half-round-trips around the ring (or the edge) doubles up the $2\tx 2$ Hamiltonian above into a $4 \tx 4$ one. Through mixing these four orbitals, we wish to find the level crossings and the flattenings. The half-round-trip introduces a $e^{i\frac{\pi}{2} (1-s)}$ factor with $s=\pm 1$. The two $s$-sectors are thus effectively separated by a $\pi$ shift in their momenta, and the on-site detuning term for the sectors is given by $s \Delta$. Paths that couple $s$-sectors with extra half-cell displacements, effectively shift the corresponding momenta by $\pi/2$. Taking all these into account we eventually arrive at,
\begin{equation}
\label{eq:AQHAQH}
  H_{\text{AQH}^\text{AQH}}\left(\vec{k}\right)=\left(
  \begin{array}{cccc}
   \alpha H_{\text{AQH}}\left(\vec{k}\right) + \Delta \mathbf{1}  & \alpha \chi\left(\vec{k}\right) \\
   \alpha \chi^\dagger \left(\vec{k}\right) & \alpha H_{\text{AQH}}\left(\vec{k} +\vec{\pi}\right) - \Delta\mathbf{1}
  \end{array}
  \right) \, ,
\end{equation}
with
\begin{equation}
    \chi\left(\vec{k}\right)=\left(
  \begin{array}{cccc}
     i\left[ d_0\left(k_x,k_y+\frac{\pi }{2}\right) + d_z\left(k_x,k_y+\frac{\pi }{2}\right) \right] & d_x\left(k_x,k_y+\pi \right)-i
     d_y\left(k_x,k_y-\pi \right)   \\
     d_x\left(k_x-\pi
     ,k_y\right)+i d_y\left(k_x+\pi ,k_y\right) & i \left[ d_0\left(k_x+\frac{\pi }{2},k_y\right) -d_z\left(k_x+\frac{\pi }{2},k_y\right) \right] 
  \end{array}
  \right) \, .
\end{equation}

The band structure of the resulting Hamiltonian, $H_{\text{AQH}^\text{AQH}}$, generated for a range of $\alpha$ is shown in Fig.~\ref{Fig:theory}. The first two bands cross at $\alpha \approx 0.25$ and flatten at $\alpha \approx 0.35$ and then cross with the two farther bands at $\alpha \approx 0.5$. The corresponding gaps are topological with $|C|=1$. These bands, therefore, successfully describe the $\text{edge}^\text{edge}$ and $\text{edge}^\text{bulk}$ states described in the main text.

\begin{figure*}[t]
    \centering
    \includegraphics[width=0.5
    \textwidth]{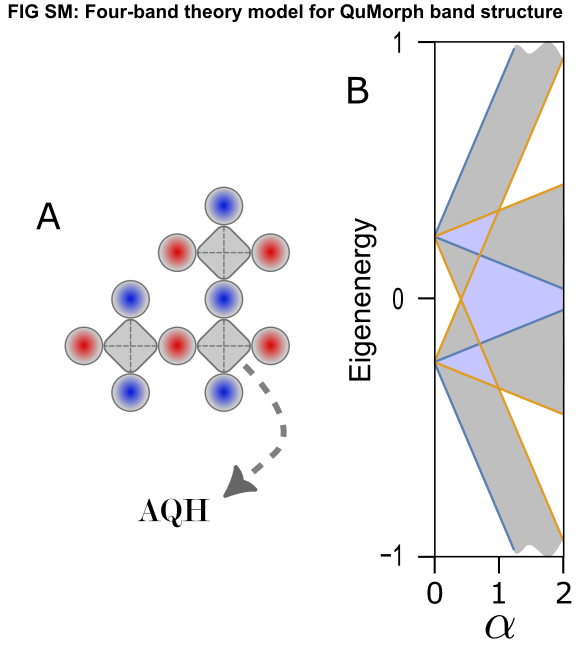}
    \caption{Four-band nested Haldane Hamiltonian description of QuMorph in HNLs. (A) Models the $\text{AQH}^\text{AQH}$ lattice with the sites of the base lattice replaced by ring resonators instead of complete AQH lattices. This model is therefore describing the $\text{edge}^\text{AQH}$ subspace. Red and blue denote the two different sublattices of the outer AQH lattice. The ring resonators are coupled with AQH hopping amplitudes proportional to $\alpha$. The dashed lines designate the next neighbor hopping while the solid lines carry a phase. The band structure of the system in (A) is shown in (B) for various values of $\alpha$.
    }
    \label{Fig:theory}
\end{figure*}
%
The expanded version of the $\text{AQH}^\text{AQH}$ model, $H_{\text{AQH}^\text{AQH}}$, is explicitly written below.
\newpage
\noindent
\begin{minipage}[t]{0.45\textwidth}
\begin{equation}
\label{eq:left-matrix}
\adjustbox{angle=90,scale=.8}{$
  \alpha \left(
  \begin{array}{cccc}
   d_z\left(k_x,k_y\right)+d_0\left(k_x,k_y\right)+\frac{\Delta }{\alpha} & d_x\left(k_x,k_y\right)-i d_y\left(k_x,k_y\right) & i
     \left(d_z\left(k_x,k_y+\frac{\pi }{2}\right)+d_0\left(k_x,k_y+\frac{\pi }{2}\right)\right) & d_x\left(k_x,k_y+\pi \right)-i
     d_y\left(k_x,k_y-\pi \right) \\
   d_x\left(k_x,k_y\right)+i d_y\left(k_x,k_y\right) & -d_z\left(k_x,k_y\right)+d_0\left(k_x,k_y\right)+\frac{\Delta }{\alpha} & d_x\left(k_x-\pi
     ,k_y\right)+i d_y\left(k_x+\pi ,k_y\right) & i \left(d_0\left(k_x+\frac{\pi }{2},k_y\right)-d_z\left(k_x+\frac{\pi }{2},k_y\right)\right) \\
   -i \left(d_z\left(k_x,k_y+\frac{\pi }{2}\right)+d_0\left(k_x,k_y+\frac{\pi }{2}\right)\right) & d_x\left(k_x-\pi ,k_y\right)-i d_y\left(k_x+\pi
     ,k_y\right) & d_z\left(k_x+\pi ,k_y+\pi \right)+d_0\left(k_x+\pi ,k_y+\pi \right)-\frac{\Delta }{\alpha} & d_x\left(k_x+\pi ,k_y+\pi \right)-i
     d_y\left(k_x+\pi ,k_y+\pi \right) \\
   d_x\left(k_x,k_y+\pi \right)+i d_y\left(k_x,k_y-\pi \right) & -i \left(d_0\left(k_x+\frac{\pi }{2},k_y\right)-d_z\left(k_x+\frac{\pi
     }{2},k_y\right)\right) & d_x\left(k_x+\pi ,k_y+\pi \right)+i d_y\left(k_x+\pi ,k_y+\pi \right) & -d_z\left(k_x+\pi ,k_y+\pi
     \right)+d_0\left(k_x+\pi ,k_y+\pi \right)-\frac{\Delta }{\alpha} \\
  \end{array}
  \right)
$}
\end{equation}
\end{minipage}\hfill
\begin{minipage}[t]{0.45\textwidth}
\begin{equation}
\label{eq:right-matrix}
\adjustbox{angle=90,scale=.5}{$
  \alpha\left(
  \begin{array}{cccc}
   \frac{\Delta }{\alpha}+2 \cos \left(k_y\right) & 2 \sqrt{2} \cos \left(\frac{k_x}{2}\right) \cos \left(\frac{k_y}{2}\right)+2 i \sqrt{2} \sin
     \left(\frac{k_x}{2}\right) \sin \left(\frac{k_y}{2}\right) & -2 i \sin \left(k_y\right) & 2 \sqrt{2} \cos \left(\frac{k_x}{2}\right) \cos
     \left(\frac{1}{2} \left(k_y+\pi \right)\right)+2 i \sqrt{2} \sin \left(\frac{k_x}{2}\right) \sin \left(\frac{1}{2} \left(k_y-\pi
     \right)\right) \\
   2 \sqrt{2} \cos \left(\frac{k_x}{2}\right) \cos \left(\frac{k_y}{2}\right)-2 i \sqrt{2} \sin \left(\frac{k_x}{2}\right) \sin
     \left(\frac{k_y}{2}\right) & \frac{\Delta }{\alpha}+2 \cos \left(k_x\right) & 2 \sqrt{2} \cos \left(\frac{1}{2} \left(k_x-\pi \right)\right) \cos
     \left(\frac{k_y}{2}\right)-2 i \sqrt{2} \sin \left(\frac{1}{2} \left(k_x+\pi \right)\right) \sin \left(\frac{k_y}{2}\right) & -2 i \sin
     \left(k_x\right) \\
   2 i \sin \left(k_y\right) & 2 \sqrt{2} \cos \left(\frac{1}{2} \left(k_x-\pi \right)\right) \cos \left(\frac{k_y}{2}\right)+2 i \sqrt{2} \sin
     \left(\frac{1}{2} \left(k_x+\pi \right)\right) \sin \left(\frac{k_y}{2}\right) & -\frac{\Delta }{\alpha}-2 \cos \left(k_y\right) & 2 \sqrt{2} \cos
     \left(\frac{1}{2} \left(k_x+\pi \right)\right) \cos \left(\frac{1}{2} \left(k_y+\pi \right)\right)+2 i \sqrt{2} \sin \left(\frac{1}{2}
     \left(k_x+\pi \right)\right) \sin \left(\frac{1}{2} \left(k_y+\pi \right)\right) \\
   2 \sqrt{2} \cos \left(\frac{k_x}{2}\right) \cos \left(\frac{1}{2} \left(k_y+\pi \right)\right)-2 i \sqrt{2} \sin \left(\frac{k_x}{2}\right)
     \sin \left(\frac{1}{2} \left(k_y-\pi \right)\right) & 2 i \sin \left(k_x\right) & 2 \sqrt{2} \cos \left(\frac{1}{2} \left(k_x+\pi
     \right)\right) \cos \left(\frac{1}{2} \left(k_y+\pi \right)\right)-2 i \sqrt{2} \sin \left(\frac{1}{2} \left(k_x+\pi \right)\right) \sin
     \left(\frac{1}{2} \left(k_y+\pi \right)\right) & -\frac{\Delta }{\alpha}-2 \cos \left(k_x\right) \\
  \end{array}
  \right)
$}
\end{equation}
\end{minipage}

\section{First-order nested topological models and lattices studied with coupled rings}\label{sm:othertopo}

Following the earlier investigation of topological models in multiple photonic systems~\cite{wang2009observation,khanikaev2013photonic,rechtsman2013photonic}, such effects have been explored in integrated ring-based systems. Table~\ref{tab:toporings} summarizes some of the proposed/realized topological models and lattices studied in coupled-ring arrays. A diverse family of HNLs with multiple configurations can be realized using these first-order nested lattices and models as the building blocks.
\begin{table}[h]
    \centering
    \begin{tabular}{|c|c|c|}
        \hline
         & Topological Model/Lattice & Reference \\
        \hline
        1    & Anomalous Quantum Hall     & Reference [\cite{mittal2019photonicaqh}]     \\
        \hline
        2    & Integer Quantum Hall     & Reference [\cite{Hafezi2013}]     \\
        \hline
        3    & Spin Hall     & Reference [\cite{yang2021optically}]     \\
        \hline
        4    & Valley Hall     & Reference [\cite{yang2021optically}]     \\
        \hline
        6    & Non-Hermitian     & Reference [\cite{hashemi2025reconfigurable}]     \\
        \hline
        7    & Hyperbolic     & Reference [\cite{huang2024hyperbolic}]     \\
        \hline
        8    & Floquet     & Reference [\cite{hashemi2024floquet}]     \\
        \hline
        9    & Higher-Order     & Reference [\cite{mittal2019photonic}]     \\
        \hline
    \end{tabular}
    \caption{Topological models and lattices realized with coupled ring systems}
    \label{tab:toporings}
\end{table}
%
\newpage
\section{Robustness}\label{sm:robust}

Fig.~\ref{Fig:robust} illustrates the scale-dependent topological robustness of edge states in HNLs against sharp corners of the square lattice.
\begin{figure*}[t]
    \centering
    \includegraphics[width=0.8
    \textwidth]{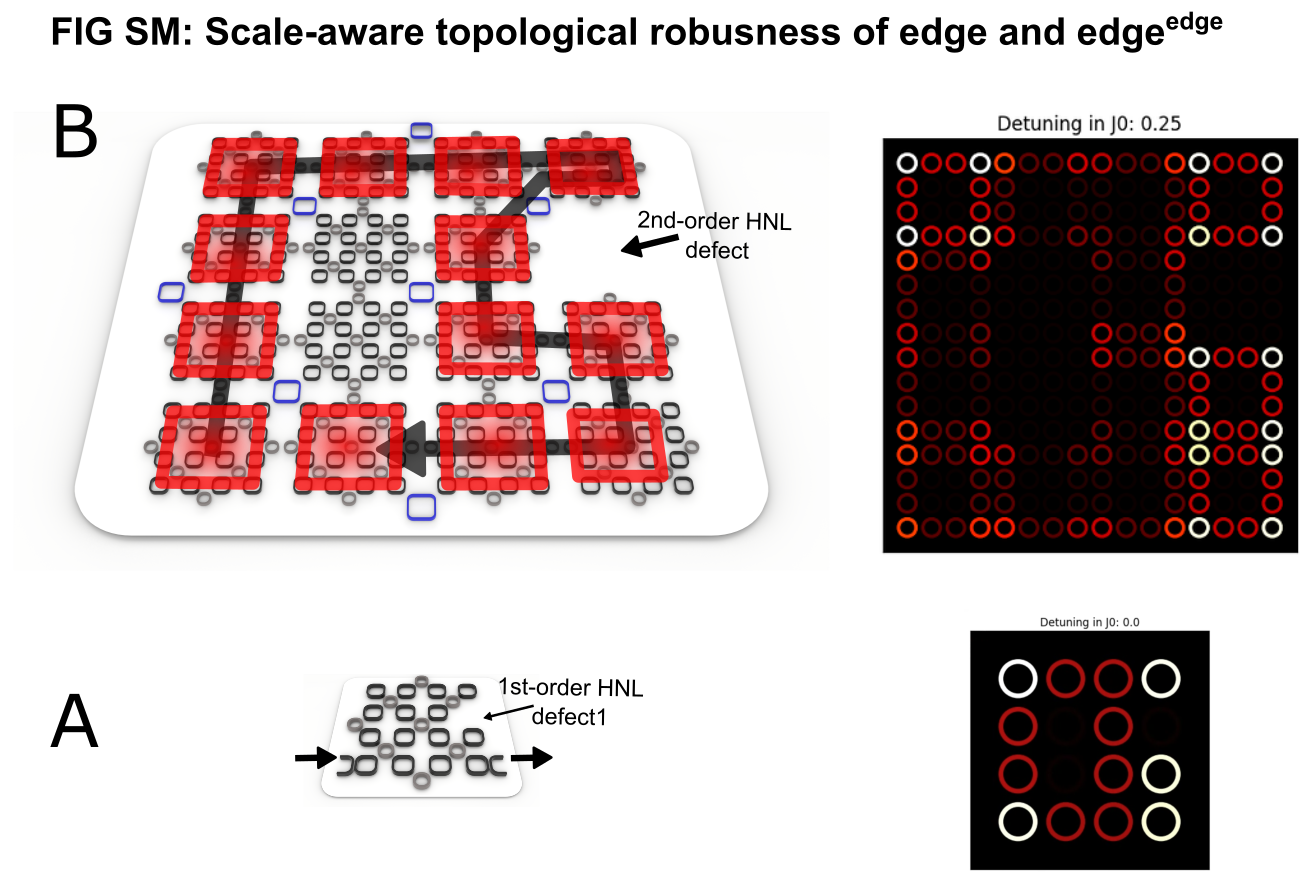}
    \caption{Robustness of HNLs. Robust edge propagation against (A) first-order and (B) 2nd order HNL defects. 
    }
    \label{Fig:robust}
\end{figure*}
%
\newpage
\section{Scale up}\label{sm:scale}

Fig.~\ref{Fig:scale} demonstrates the bottom-up construction of the HNLs from single rings to single plaquettes and second-order nested HNLs with different dimensionalities.
\begin{figure*}[t]
    \centering
    \includegraphics[width=0.8
    \textwidth]{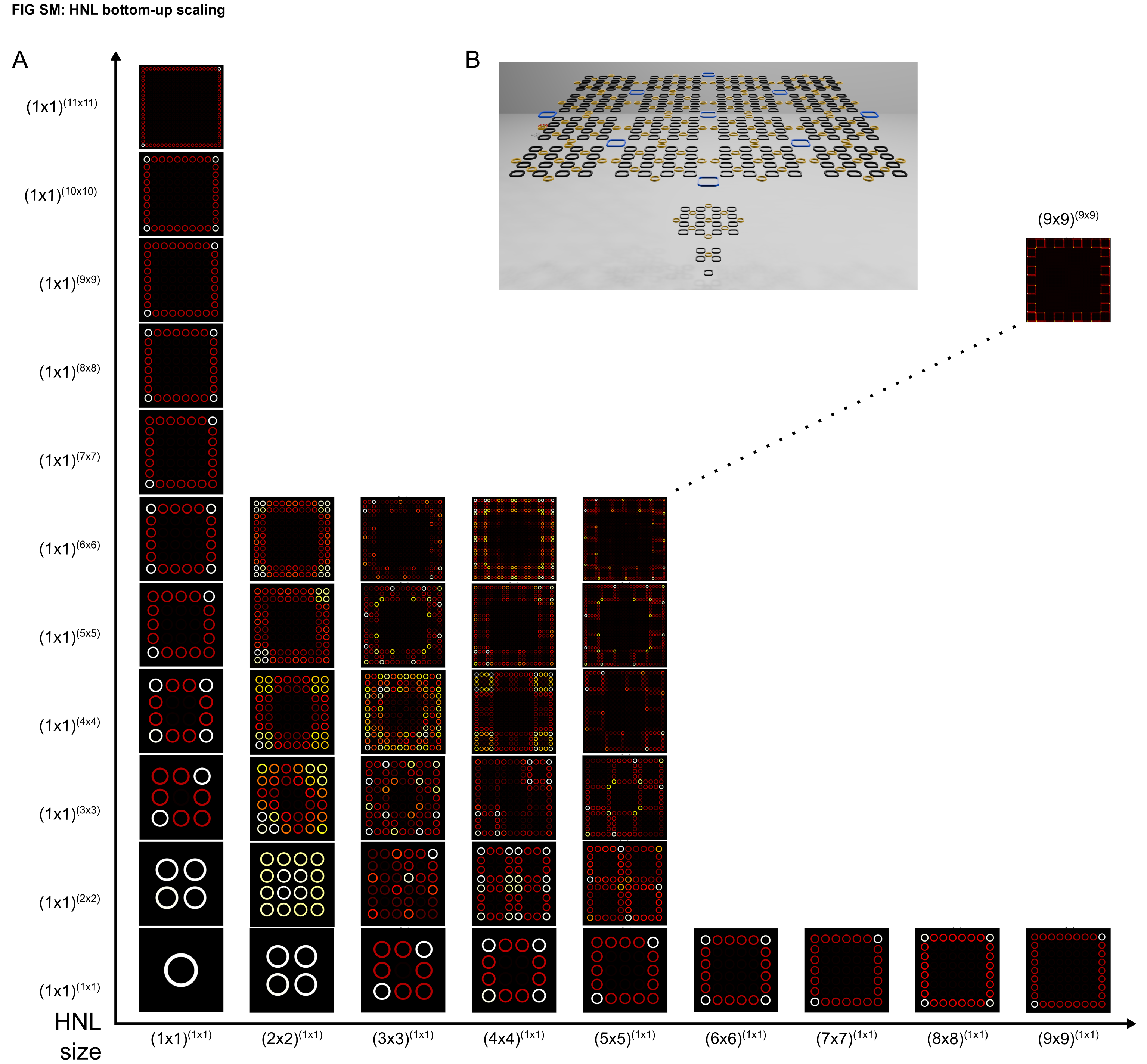}
    \caption{(A) Bottom-up scaling of HNLs. For each HNL size, a typical edge or $\text{edge}^\text{edge}$ profile is shown. (B) Schematic of the scale-up. 
    }
    \label{Fig:scale}
\end{figure*}

\newpage
\section{Eigenvalues versus drop-port spectra in AQH-AQH HNLs}\label{sm:eigen}

Fig.~\ref{Fig:eigendropaqh} illustrates the QuMorph of the eigenvalues in HNLs, and the corresponding frequency nesting probed at the drop-port of the devices.
\begin{figure*}[t]
    \centering
    \includegraphics[width=0.8
    \textwidth]{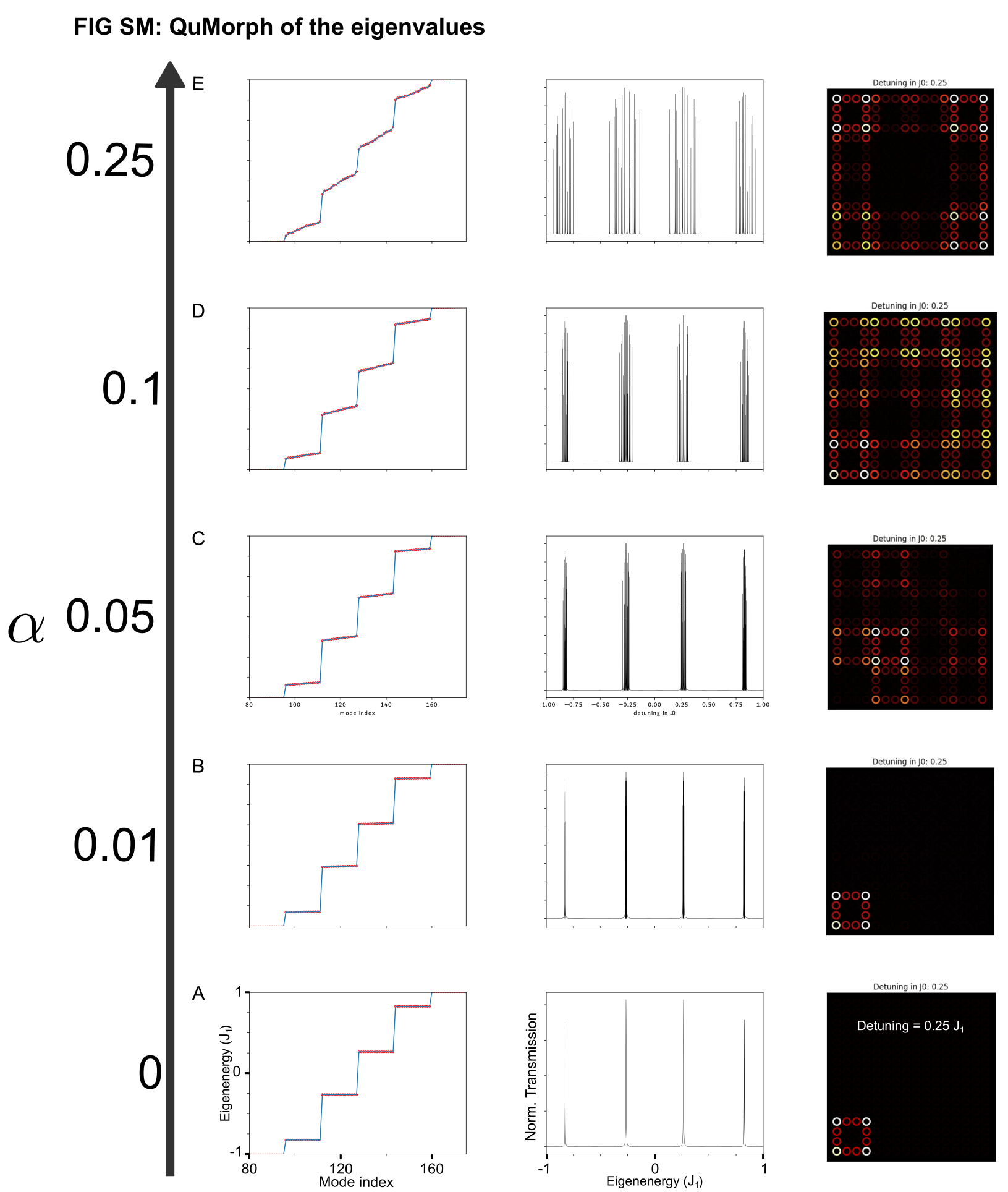}
    \caption{(A-E) Eigenvalues, drop-port transmission spectra, and intensity field profiles (at the fixed pump frequency $0.25 J_1$) in $4\tx 4^{4\tx 4}$ $\text{AQH}^\text{AQH}$ HNL, as a function of $\alpha$.  
    }
    \label{Fig:eigendropaqh}
\end{figure*}

\section{Eigenvalues versus drop-port spectra in mixed-topology HNLs}\label{sm:eigenmixed}

Fig.~\ref{Fig:eigendropaqhmix} illustrates the eigenvalues, drop-port transmission spectra, and intensity field profiles in HNLs with mixed-topology of different configurations.
\begin{figure*}[t]
    \centering
    \includegraphics[width=0.8
    \textwidth]{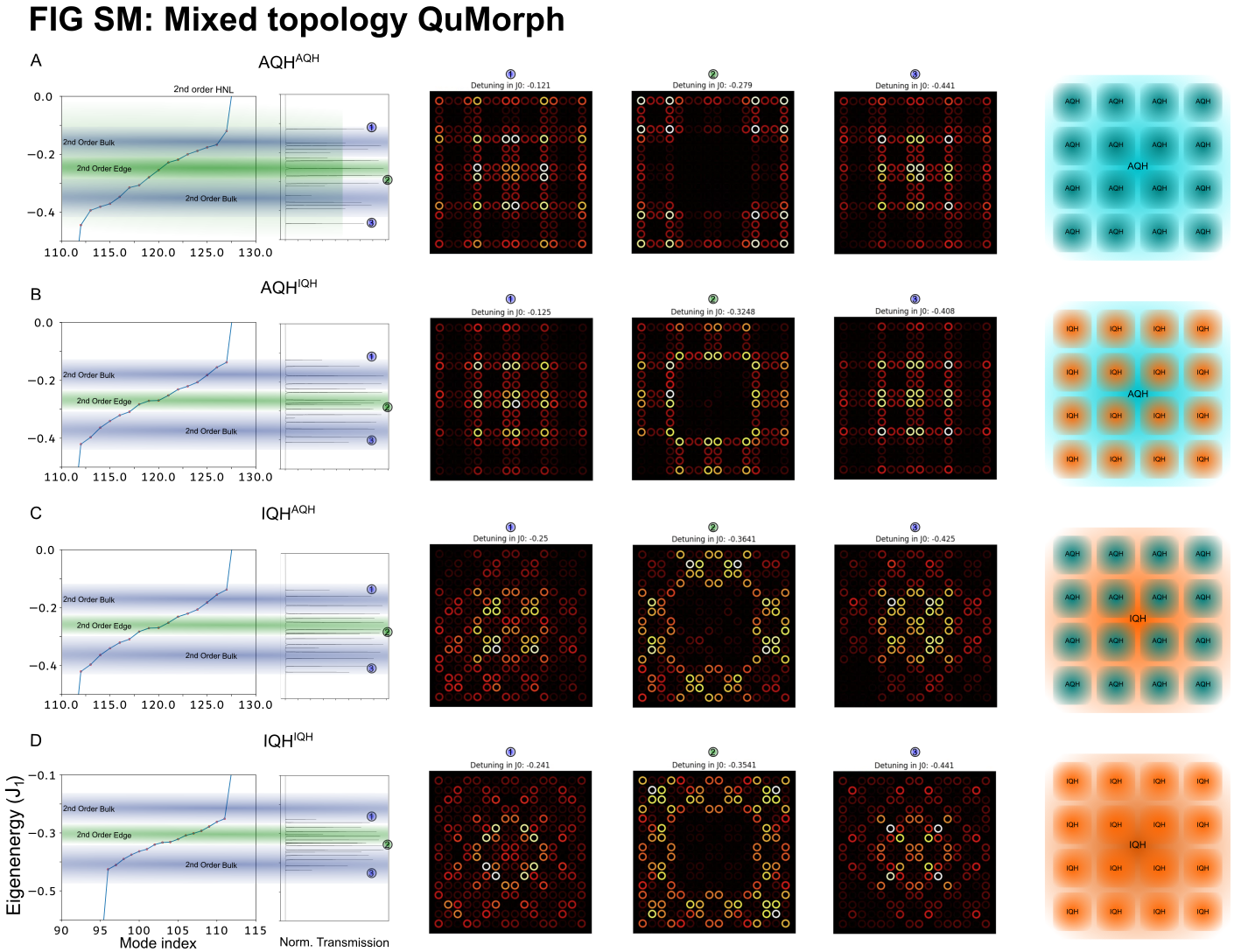}
    \caption{(A-D) Eigenvalues, drop-port transmission spectra, and intensity field profiles in a $4\tx 4^{4\tx 4}$ HNL with $\text{AQH}^\text{AQH}$, $\text{AQH}^\text{IQH}$, $\text{IQH}^\text{AQH}$, and $\text{IQH}^\text{IQH}$ topologies, respectively.
    }
    \label{Fig:eigendropaqhmix}
\end{figure*}

\newpage
\section{Complete sets of field profiles}\label{sm:complete}

Fig.~\ref{Fig:full} illustrates nesting of the eigenvalues, drop-port transmission spectra, and intensity field profiles in a single-layer nested AQH lattice compared with its HNL counterpart.
\begin{figure*}[t]
    \centering
    \includegraphics[width=0.95
    \textwidth]{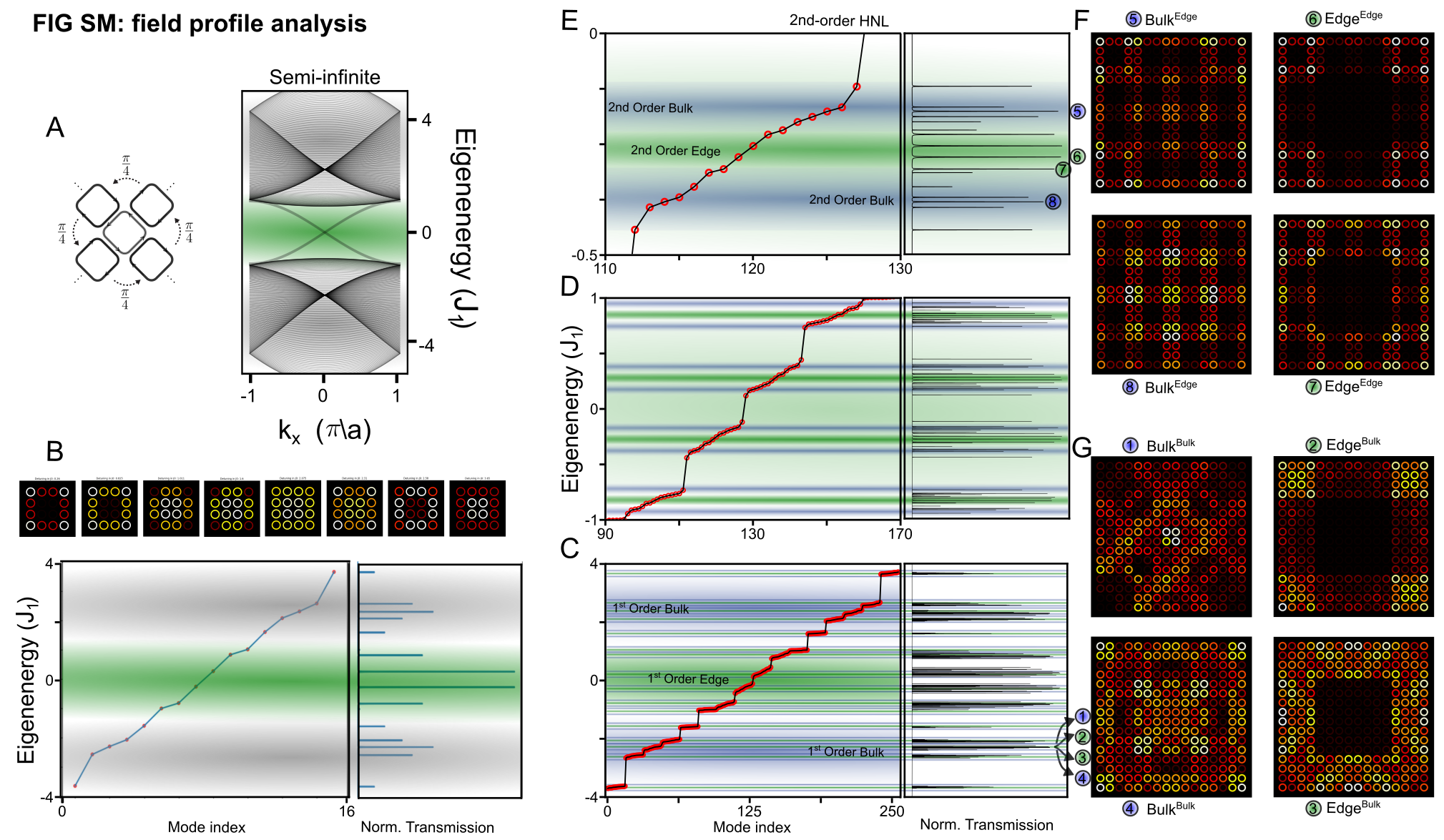}
    \caption{(A) Unit cell and the band structure of a semi-infinite AQH lattice. (B) Eigenvalues and the drop-port spectra of a $4 \tx 4$ AQH lattice. (E-F) Eigenvalues, drop-port transmission spectra, intensity field profiles of a $4\tx 4^{4\tx 4}$ $\text{AQH}^\text{AQH}$ HNL.  
    }
    \label{Fig:full}
\end{figure*}

\newpage
\section{QuMorph at large $\alpha$}\label{sm:large}

Fig.~\ref{Fig:large} illustrates the drop-port transmission spectra of an HNL at larger values of $\alpha$.
\begin{figure*}[t]
    \centering
    \includegraphics[width=0.6
    \textwidth]{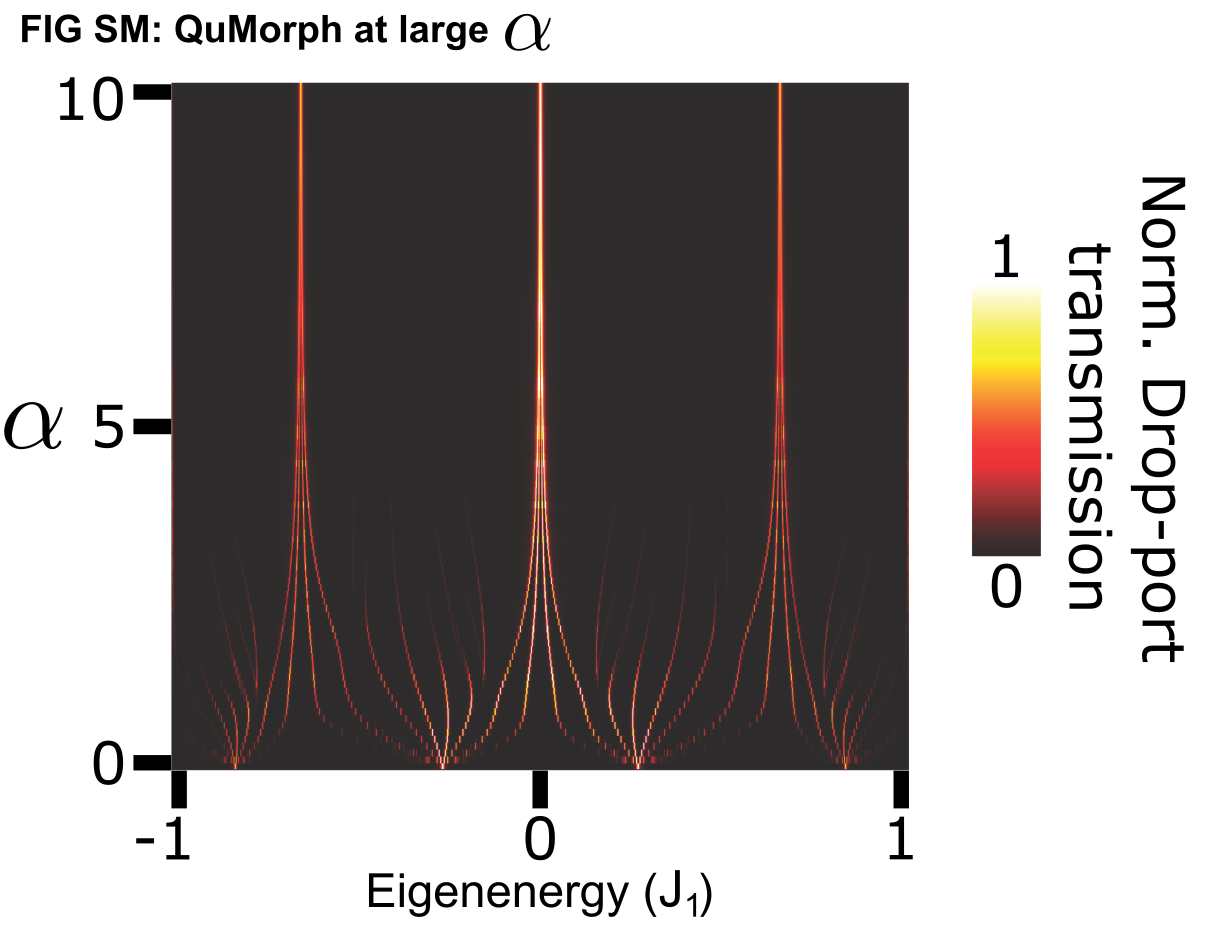}
    \caption{Drop-port transmission spectra of a $4\tx 4^{4\tx 4}$ $\text{AQH}^\text{AQH}$ HNL as a function of $\alpha$. 
    }
    \label{Fig:large}
\end{figure*}

\newpage
\section{Tunable Fractal Dimension}\label{sm:fract}
\begin{figure*}[h]
    \centering
    \includegraphics[width=0.8
    \textwidth]{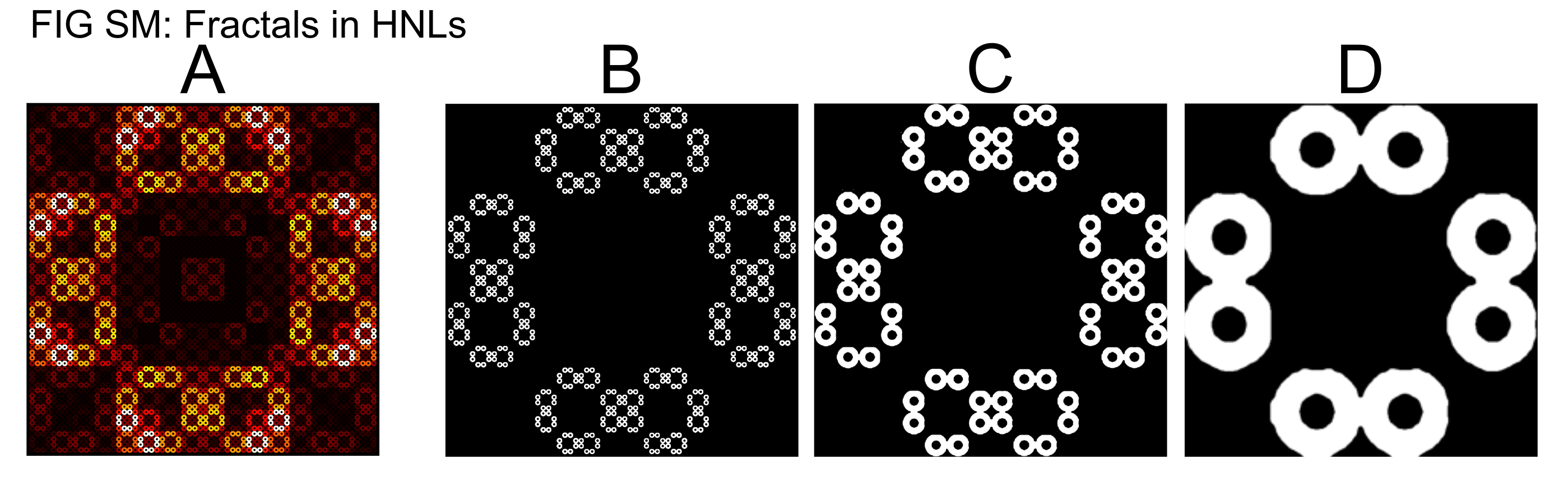}
    \caption{Fractals naturally appear in HNLs. (A) is the wave-function profile of a $4\tx 4^{4\tx 4^{4\tx 4}}$ $\text{AQH}^{\text{AQH}^{\text{AQH}}}$ lattice at $E\approx2.6 J_1$. (B) Depicts the filled/empty version of the original wave-function. (B) A self-similar fractal, with its constituents in (C) and (D) repeating its shape. The wave-function depicted above has, therefore, a fractal dimension of $d=3/2$.
    }
    \label{Fig:fractal}
\end{figure*}
We investigate fractals and their dimensions which appear naturally in HNLs, as summarized in Fig.~\ref{Fig:fractal}. To avoid technicalities, imagine that we have a hierarchical AQH system of infinite order; an AQH lattice nested in another, nested in another, and so on. In this system we can choose the energy range, as explained in Sec.~\ref{sec:BulkEdge}, to lie where only the edge states of all orders are activated, so that what we have is the $\text{edge}^{\text{edge}^{.^{.^.}}}$ state. In that instance, the wave-function configuration has a fractal dimension of one, $D=1$, or, depending on the lattice sizes and their designs, slightly less than one, $D\approx 1$. However, we can also tune the system so that the bulk states are activated in all orders to achieve a $\text{bulk}^{\text{bulk}^{.^{.^.}}}$ state. In that situation, the wave-function configuration has a fractal dimension of two, or almost two, depending on the design and size of the lattices, $D \approx 2$. Since we can continuously change the detuning energy and the tuning parameter going from one to the other state we expect to observe configurations with fractal dimensions in the range $1\leq D \leq 2$.

We can define and determine the fractal dimension in a probabilistic fashion, however for simplicity, let us employ the Minkowski–Bouligand method in order to derive the fractal dimension~\cite{FractalMattila,FractalFalconer,FractalBarral}: First we divide rings into filled and empty rings, by looking at the normalized power distributed on each ring. Then we calculate the fractal dimension of the filled ring configuration by going up and down the hierarchy ladder. Since the density, $\rho=M/L^d$, of a self-similar object should not change at any step on the ladder, we can write,
\begin{equation}
    \frac{\rho}{\rho}= \frac{M_H/M_L}{L_H^d/L_L^d} =1 \, ,
\end{equation}
where $M$ is the ``mass'' or here the number of filled rings and $L$ is the size of the system which for example can be the number of rings that fit on one side of the hierarchical lattice. $H$/$L$ subscripts designate higher and lower steps of the ladder, and $d$ is the fractal dimension of the configuration, which from above is given by,
\begin{equation}
    d= \frac{\ln M_H/M_L}{\ln L_H/L_L} \, .
\end{equation}
It should be noted that one can not provide real examples of a hierarchy of infinite order, but the method above works well for HNLs with finite nesting orders. Fig.~\ref{Fig:fractal} shows an example of a fractal in the third-order HNL. A few other instances are provided in Fig.~\ref{Fig:fractalist}.
\begin{figure*}[h]
    \centering
    \includegraphics[width=0.8
    \textwidth]{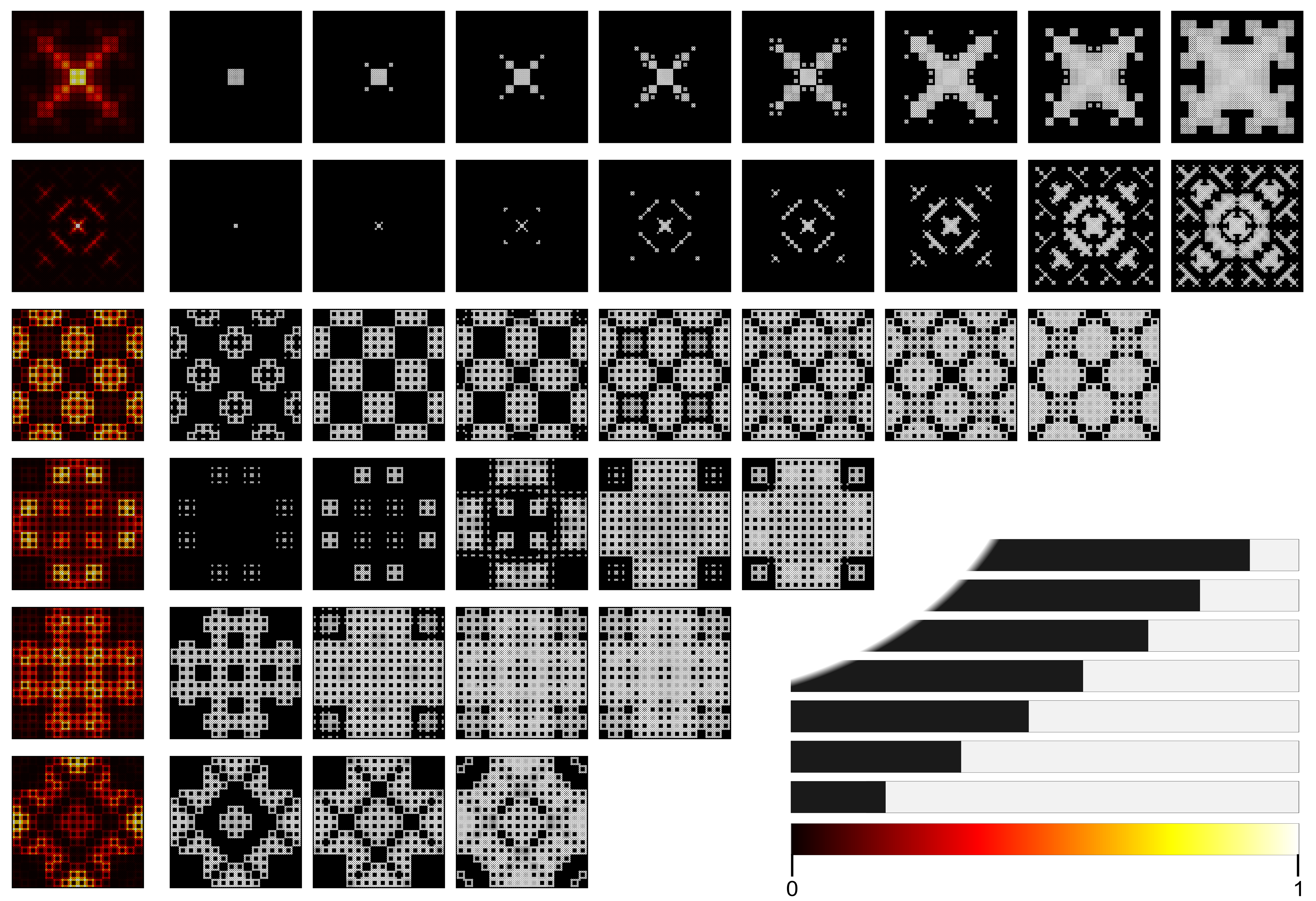}
    \caption{Fractal patterns appear in the $4\tx 4^{4\tx 4^{4\tx 4}}$ $\text{AQH}^{\text{AQH}^{\text{AQH}}}$ HNL lattice. Rows: field profile examples, alongside their empty-filled plots on their right, at various cut-offs. The cut-off color bar is schematic.
    }
    \label{Fig:fractalist}
\end{figure*}

\section{Lattice Hamiltonian}\label{sm:latham}
We nest one Hamiltonian, $H^i$, into another, $H^f$, by going to the spatial basis and setting a link up between the copies of $H^i$ according to $H^f$. (Note the change of super to subscription compared to the main text for the sake of simplicity.) Let $a^\dagger_{{i,j}^{n,m}}$ be the creation operator that points to the $(i,j)$ position inside of the $(n,m)$ copy of $H^i$. Then
\begin{equation}
   H_{nm}^i = \sum_{ij} \left( H^0_{ij} + J_i \sum_{\langle\langle i'j' \rangle\rangle}   e^{i\phi_{ij,i'j'}} a^\dagger_{{ij}^{nm}} a_{{i'j'}^{nm}} \right) \, ,
\end{equation}
where $H^0_{i,j}$ is the on-site potential at position $(i,j)$ inside the inner lattice, $J_i$ the hopping rate between $H^i$ sites with appropriate phases, $\phi$, describing an inner AQH lattice at the position $(n,m)$ inside the outer lattice. The $\text{AQH}^\text{AQH}$ HNL is given by similarly connecting the $H_i$ copies,
\begin{equation}
    H= \sum_{nm} \left( H^i_{nm} + J_f \sum_{\langle\langle n'm' \rangle\rangle}   e^{i\phi_{nm,n'm'}} a^\dagger_{{c_{nm}}^{nm}} a_{{c_{n'm'}}^{n'm'}} \right) \, .
\end{equation}
The $c_{nm}$ indices determine which parts of the $H_i$ copies are connected within the nested system. In our design (Fig.~\ref{Fig:implement}) $c_{nm}$ point to the corners of $H_i$: top-right and bottom-left when $n+m$ is odd and to the other two corners otherwise.
Noticing how $H^i_{nm}$ sits within the definition of $H$, the generalization to higher nesting orders becomes straightforward. Yet the most straightforward way to describe a lattice Hamiltonian is to draw it, the content of Fig.~\ref{Fig:HNL}
\begin{figure*}[h]
    \centering
    \includegraphics[width=0.8
    \textwidth]{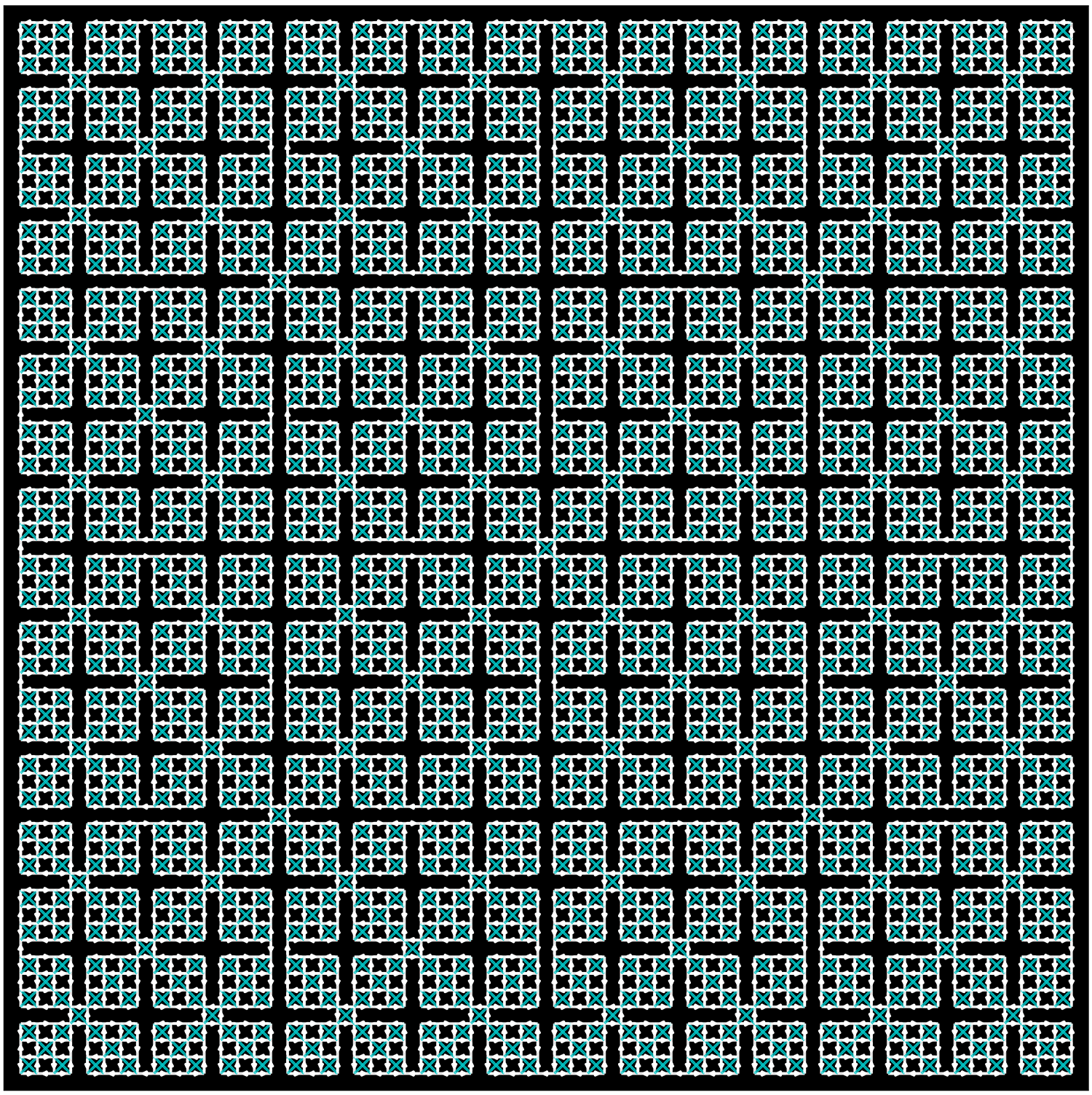}
    \caption{The $4\tx 4^{4\tx 4^{4\tx 4}}$ $\text{AQH}^{\text{AQH}^{\text{AQH}}}$ HNL lattice. The white arrows carry a phase of $e^{i\pi/4}$. Blue denotes next neighbor hoppings of the AQH lattices.
    }
    \label{Fig:HNL}
\end{figure*}

\section{Nonlinear simulations in HNLs}\label{sm:nonl}
For the QuMorph nonlinear simulations presented in the main text, in the $\text{AQH}^\text{AQH}$ HNL with $4\tx 4^{4\tx 4}$ dimensions and $\alpha = 0.3$, we closely follow the LLE formalism, normalization method, and nonlinear simulation parameter regimes described in the SI of Ref.~\cite{mittal2021topological}. In the following, we highlight the key considerations and simulation parameters. It should be noted that due to the large lattice size, the three-timescale nature, and the high spectral resolution required to resolve the hyper-nested comb teeth structures, these simulations are demanding on both computational speed and data storage size. We perform these simulations on workstations equipped with 128 GB of RAM, 3.2 GHz CPU cores, and multiple terabytes of hard drive storage. To observe the soliton formation, we set the spectral resolution for the pump frequency sweep to 0.00002~$J_1$, and the number of iterations to 10000 steps with a step size of~$0.1/J_1$. We note that the pump frequency sweep is performed over a range of 0.281~$J_1$ to 0.274~$J_1$, over a cold cavity resonance that corresponds to the slowest timescale ($\tau_2$) of the HNL. Furthermore, we note that this pump sweep range is much smaller than the slowest-timescale mode spacing of the HNL ($1/\tau_2$). Several sweeps are performed to identify the OPO threshold, using the simulation parameters that follow: the second order (anomalous-regime) dispersion $D_2$ = 0.001~$J_1$, the extrinsic/intrinsic coupling/loss rates $\kappa_{ex}=0.01 J_1$ and $\kappa_{in}=0.0001 J_1$, respectively. Following the normalization method in Ref.~\cite{mittal2021topological}, with the nonlinear interaction strength $\gamma$ and the pump field amplitude $\mathcal{E}$, we set the normalized pump field $F=\sqrt{2\kappa_{ex}\gamma/J_1^3}\mathcal{E}$ to 0.036 to access the DKS regime studied in the main text. For the simulation parameters, we set the total number of longitudinal modes of the single ring $\mu$ to 257, ranging from $\mu = $ -128 to $\mu = $  128, with the pump mode being $\mu = $ 0.
\end{supplement}
\bibliography{Main.bib}
\end{document}